\documentclass[oldversion]{aa}
\usepackage{natbib}
\bibpunct{(}{)}{;}{a}{}{,}
\usepackage{graphicx}
\usepackage{txfonts}
\usepackage{rotating}
\begin{document}

\title{Spectral atlas of massive stars around \ion{He}{i}~10830~{\AA}\thanks{Based on
observations made at Observat\'orio do Pico dos Dias/LNA (Brazil)}}

\author{J. H. Groh \inst{1} \and  A. Damineli \inst{1} \and F. Jablonski \inst{2}}
\institute{Instituto de Astronomia, Geof\'{i}sica e Ci\^encias Atmosf\'ericas,
Universidade de S\~ao Paulo, Rua do Mat\~ao 1226, Cidade Universit\'aria,
05508-900, S\~ao Paulo, SP, Brasil \and Instituto Nacional de Pesquisas Espaciais/MCT, Avenida dos
Astronautas 1758, 12227-010 S\~ao Jos\'e dos Campos, SP, Brasil}

\authorrunning{Groh, Damineli \& Jablonski}
\titlerunning{Atlas of massive stars around \ion{He}{i}~10830 {\AA}}

\offprints{Jose Henrique Groh, \email{groh@astro.iag.usp.br}}
\abstract{We present a digital atlas of peculiar, high-luminosity massive stars in the
near-infrared region (10470--11000~{\AA}) at medium resolution ($R\simeq7000$). The spectra
are centered around \ion{He}{i}~10830~{\AA}, which is formed in the wind of those stars, and is
a crucial line to obtain their physical parameters. The instrumental configuration also
sampled a rich variety of emission lines of \ion{Fe}{ii}, \ion{Mg}{ii}, \ion{C}{i},
\ion{N}{i}, and Pa\,$\gamma$. Secure identifications for most spectral lines are given,
based on synthetic atmosphere models calculated by our group. We also propose that two
unidentified absorption features have interstellar and/or circumstellar origin.
For the strongest one (10780~{\AA}) an empirical calibration between E(B-V) and equivalent
width is provided. The atlas displays the spectra of massive stars organized in four
categories, namely Be stars, OBA Iape (or luminous blue variables, LBV candidates and
ex/dormant LBVs), OB supergiants and Wolf-Rayet stars. For comparison, the photospheric
spectra of non emission-line stars are presented. Selected LBVs were observed in different
epochs from 2001 to 2004, and their spectral variability reveals that some stars, such as
$\eta$ Car, AG Car and HR Car, suffered dramatic spectroscopic changes during this time
interval. }
\maketitle
\section{Introduction\label{intro}}

In the last decades tremendous progress has been achieved in understanding how a
massive star evolves along the HR diagram. A new scenario of massive star evolution was
developed after the models allowed for the effects of rotation in the stellar structure
and evolution \citep{maeder_araa00}. Also, the physical parameters of massive stars can
currently be constrained  with the advent of a new generation of
fully-blanketed, spherical-symmetric, non-LTE radiative transfer codes
\citep{hubeny95,hm98,hamann98,pauldrach01,grafener02,puls05}. Even for extremely massive objects such as Eta Carinae
\citep{hillier01} we now have a much better comprehension of the physical
parameters of the central star by analyzing the emerged spectrum.

However, the short transitional stages of massive star evolution are
still poorly understood. The radiation emitted by the photosphere of these stars 
interacts with the optically-thick, dense wind, which makes the spectral classification
a very challenging task. In addition to the luminosity, effective
temperature and surface gravity, several other stellar parameters might change
the spectral morphology of these objects, notably the mass-loss rate and the wind terminal
velocity. In the discussion on the characteristics of the atlas, we point out the similarities and
differences in the spectral morphology of massive stars belonging to different evolutionary stages.
\defcitealias{wf2000}{WF2000}

The aim of this paper is to extend the idea of the {\it OB Zoo} published by \citet{wf2000}
(hereafter WF2000) to the near-infrared region of the spectrum. 
Some of the objects
present in this work were observed at lower resolution from 1.0 to 2.5 $\mu$m by
\citet{mcgregor88}. In addition, peculiar massive stars were also observed 
around 8500 {\AA} and 10000 {\AA} by \citet{lopes92}, and in the H and K bands
by \citet{morris96}. 

We chose to explore the ``forgotten" region around
\ion{He}{i}~10830~{\AA} due to the strength of this line,
 which is often the strongest one in
the near-infrared region. Pioneering
observations of early-type stars around this line were published by
\citet{andrillat79} and by \citet{meisel82}.
Especially for objects with dense winds and high terminal
velocities, \ion{He}{i}~10830~{\AA} is one of the few strong He~I lines that is unblended. This line is
formed over a large region of the wind and thus it is a crucial diagnostic of the physical
properties of the wind and of the underlying photosphere. Moreover, with a relatively short
spectral coverage around 10830\,{\AA}, we can sample a number of strong, unblended lines,
such as Pa\,$\gamma$, and rich emission-line spectra of \ion{Fe}{ii}, \ion{Mg}{ii}, \ion{C}{i},
and \ion{N}{i}. They provide important constraints on the physical conditions found in
those stars and their winds.

We observed other peculiar objects that are not present in \citetalias{wf2000}, but are as
fascinating and intriguing as their original program stars. Meanwhile, we did not observe
some objects present in \citetalias{wf2000}, especially the faintest ones, that were beyond
the limiting magnitude of our instrumental configuration. Nevertheless, we tried to cover a large
range of evolutionary phases
and physical parameters found in hot stars. However, it is beyond the scope of this paper
to present a full evolutionary sequence of massive stars around 10830\,{\AA}, especially considering
that a significant fraction of the objects shown here does not have a
well-determined evolutionary stage.

As part of the atlas, we present the spectral variability of a large sample of luminous
blue variables (henceforth LBV) and LBV candidates during a time interval of 3 years. To
the best of our knowledge, this is the first work in any spectral region to present the
spectroscopic variability on a timescale of a few years (which is typical for S Dor-type
variations), for a large sample of LBVs and related objects. As we show along the atlas,
different behaviors of variability can be noticed (radial velocity, line profile,
strength). 

This paper is organized as follows. In Sect. \ref{obs} we describe the observations made at the Observat\'orio
Pico dos Dias (LNA/Brazil); in Sect. \ref{atlas} we present the atlas, briefly contextualizing and introducing each
class of objects, and then describing the past bibliography and the spectral features of each
individual object. In Sect. \ref{discussion} we discuss the unidentified features present in the
program stars.

\begin{table*}[h!]
\footnotesize
\caption{Basic data for the program stars, ordered by increasing right ascension. Projected
rotational velocities are indicated for Be stars, while the presence of a circumstellar nebula is indicated for LBVs and related objects.
References are: CN04=\citet{cn04},
Cro06=\citet{crowther06}, Mir01=\citet{mir01},
L97=\citet{l97}, LL06=\citet{ll06}, Riv97=\citet{rivinius97}, R97=\citet{roche97}, 
S06=\citet{s06}, Sta03=\citet{stahl03},
vG92=\citet{vg92}, vG01=\citet{vg01},
WF2000=\citet{wf2000}, Z05=\citet{zorec05} }
\label{tabledata} 
\centering
\begin{tabular}{c c c c c c c c c}
\multicolumn{9}{c}{Be stars} \\
\hline\hline
Name & HD & RA (J2000) & DEC (J2000)  & Spectral Type & $v\,sin{ \mathrm{i}}$& Reference & Fig.& Obs. Date \\
\hline
X Persei   & 24534  & 03 55 23.1 & +31 02 45.0       & B0Ve   & 293 & L97,R97,Z05  & \ref{be3} & 02Nov04  \\
$\beta$ Mon\,A   & 45725  & 06 28 49.4 & -07 02 03.5 & B3Ve   & 330 & Z05      & \ref{be3}     & 02Nov04 \\
$\delta$ Cen  & 105435 & 12 08 21.5 & -50 43 20.7    & B2Vne  & 240 & LL06     & \ref{be2}     & 01Jun09 \\
HD 110432  & 110432 & 12 42 50.3 & -63 03 31.0       & B1IVe  & 300 & S06      & \ref{be3}     & 02May01 \\
HD 120991  & 120991 & 13 53 57.2 & -47 07 41.4       & B2Ve   & 130 & LL06     & \ref{be2}     & 01Jun09  \\
$\eta$ Cen    & 127972 & 14 35 30.4 & -42 09 28.2    & B2Ve   & 310 & LL06     & \ref{be1}     & 01Jun09   \\
HD 142983  & 142983 & 15 58 11.4 & -14 16 45.7       & B3IVe  & 390 & Z05      & \ref{be1}     & 01Jun09   \\
$\delta$ Sco  & 143275 & 16 00 20.0 & -22 37 18.2    & B0.3IVe & 150 & Mir01   & \ref{be3}     & 02Apr29  \\
$\chi$ Oph    & 148184 & 16 27 01.4 & -18 27 22.5    & B1.5Ve & 144 & Z05      & \ref{be2}     & 01Jun12      \\
$\alpha$ Ara   & 158427 & 17 31 50.5 & -49 52 34.1   & B2Vne  & 270 & LL06     & \ref{be1}     & 01Jun09    \\
66 Oph     & 164284 & 18 00 15.8 & +04 22 07.0       & B2Ve   & 220 & LL06     & \ref{be1}     & 01Jun09     \\
$\lambda$ Pav & 173948 & 18 52 13.0 & -62 11 15.3    & B2IIIe  & 140 & Z05     & \ref{be3}     & 02Apr29     \\
$\upsilon$ Cyg & 202904 & 21 17 55.1 & +34 53 48.8   & B2 Ve   & 185 & Z05     & \ref{be2}     & 01Jun10  	  \\
\hline   
&&&&&\\
\end{tabular}
\begin{tabular}{c c c c c c c c c}
\multicolumn{9}{c}{LBVs, LBV candidates and ex/dormant LBVs} \\
\hline\hline
Name & HD & RA (J2000) & DEC (J2000)  & Spectral Type & Nebula  & Reference &  Fig. & Obs. Date(s) \\
\hline
HD 87643   & 87643  & 10 04 30.3 & -58 39 52.1      & Bpec	   & yes   & WF2000	 & \ref{bpec1}  	    & 01Jun12		  \\
HR Car     & 90177  & 10 22 53.8 & -59 37 28.4      & B3Ia--A2Ia    & yes  & WF2000,vG01  & \ref{lbv1}, \ref{var3}  & 01Jun12, 04Apr30   \\
$\eta$ Car & 93308  & 10 45 03.6 & -59 41 03.3        & Bpec	     & yes   & WF2000,vG01 & \ref{lbv2}, \ref{var2} & 01Jun12, 02Nov04,  \\
&&&&&&&&  03Jun28, 04Apr30 \\
GG Car     & 94878  & 10 55 58.0 & -60 23 33.4      & Bpec	     & no    & WF2000	   & \ref{bpec1}	    & 01Jun12	    \\
AG Car     & 94910  & 10 56 11.6 & -60 27 12.8   & WN11h--Apec      & yes  & WF2000,vG01  & \ref{lbv3}, \ref{var1}  & 01Jun09,02Nov04,  \\
&&&&&&&& 04Apr30, 04Jul02 \\ 
He 3-519   & -      & 10 53 59.6 & -60 26 44.3   & WN11h	    & yes  & WF2000,vG01  & \ref{lbv2}, \ref{var2}  & 01Jun12, 04Apr30  \\
W 243      & -      & 16 47 07.5 & -45 52 28.5       & A2Ia	     & no   & CN04  & \ref{lbv1}		    & 04Jul02	    \\
HD 326823  & 326823 & 17 06 53.9 & -42 36 39.7       & Bpec	     & no    & WF2000	    & \ref{bpec1}	    & 01Jun12	    \\
HD 160529  & 160529 & 17 41 59.0 & -33 30 13.7   & A2Ia 	     & no   & Sta03	   & \ref{lbv1}, \ref{var3} & 01Jun12, 04Apr30   \\
HD 316285  & 316285 & 17 48 14.0 & -28 00 53.5   & Bpec 	     & no    & WF2000	   & \ref{lbv2}, \ref{var2} & 01Jun09,02Nov04,   \\
&&&&&&&& 04Apr30 \\
HD 168607  & 168607 & 18 21 14.7 & -16 22 32.1   & B9Iape	     & no   & WF2000	   & \ref{lbv1}, \ref{var3} & 01Jun09, 04Jul02   \\
HD 168625  & 168625 & 18 21 19.5 & -16 22 26.0       & B6Iap	     & yes  & WF2000	   & \ref{lbv1}, \ref{var3} & 01Jun09		 \\
MWC 314    & -      & 19 21 34.0 & +14 52 56.9   & Bpec 	     & no    & vG01,Mir03  & \ref{lbv3}, \ref{var1} & 01Jun09,02Nov04,   \\
&&&&&&&& 04Apr30, 04Jul02 \\
P Cyg      & 193237 & 20 17 47.2 & +38 01 58.5   & Bpec 	     & yes   & WF2000,vG01 &\ref{lbv3}, \ref{var1}  & 01Jun10,02Nov04,  \\
&&&&&&&& 04Jul02\\  
\hline
&&&&&&&&\\
\end{tabular}
\begin{tabular}{c c c c c c c c}
\multicolumn{8}{c}{OB Supergiants} \\
\hline\hline
Name & HD & RA (J2000) & DEC (J2000)  & Spectral Type & Reference & Fig. & Obs. Date(s) \\
\hline
HD 80077   & 80077  & 09 15 54.8 & -49 58 24.6         & B2Ia+  &  vG92, vG01	      & \ref{obsuper}	  & 01Jun12	      \\
HD 151804  & 151804 & 16 51 33.7 & -41 13 49.9        & O8Iaf	& WF2000    & \ref{obsuper}		  & 01Jun09	       \\
WR 79a     & 152408 & 16 54 58.5 & -41 09 03.1         & O8:Iafpe,WN9ha & WF2000	  & \ref{obsuper} & 01Jun09	       \\
$\zeta^1$ Sco  & 152236 & 16 53 59.7 & -42 21 43.3 & B1.5Ia+  & Cro06	& \ref{obsuper}, \ref{var3}	  & 01Jun11, 04Jul02 	\\
HD 154090  & 154090 & 17 04 49.3 & -34 07 22.5       & B0.7Ia	& Cro06   & \ref{obsuper}		  & 01Jun09, 04Jul02 	\\
HD 169454  & 169454 & 18 25 15.1 & -13 58 41.5      & B1Ia+    & Riv97    &\ref{obsuper}, \ref{var3}	  & 01Jun09, 04Jul02 	\\
\hline
&&&&&\\
\end{tabular}
\begin{tabular}{c c c c c c c c}
\multicolumn{8}{c}{Wolf-Rayet stars} \\
\hline\hline
Name & HD & RA (J2000) & DEC (J2000) & Spectral Type & Reference & Fig. & Obs. Date(s)  \\
\hline
WR 6       & 50896  & 06 54 13.0 & -23 55 42.0  & WN4	   & vdH01    &  \ref{wn}	 & 02Nov04   \\
WR 11      & 68273  & 08 09 32.0 & -47 20 11.7  & WC8+O7.5III-V & vdH01    &  \ref{wc}   & 01Jun11   \\
WR 22      & 92740  & 10 41 17.5 & -59 40 36.9  & WN7h+O9III-V  & vdH01    &  \ref{wn}   & 01Jun09   \\
WR 78      & 151932 & 16 52 19.2 & -41 51 16.2  & WN7h    & vdH01    &  \ref{wn}	 & 01Jun09   \\
WR 90      & 156385 & 17 19 29.9 & -45 38 23.9  & WC7	   & vdH01    &  \ref{wc}	 & 01Jun11   \\
WR 103     & 164270 & 18 01 43.1 & -32 42 55.2  & WC9d     & vdH01    &  \ref{wc}	 & 01Jun11   \\
WR 136     & 192163 & 20 12 06.5 & +38 21 17.8  & WN6(h)     & vdH01	&  \ref{wn}	 & 01Jun09   \\
\hline
\end{tabular}
\end{table*}

\section{Observations\label{obs}}

The near-infrared spectra were obtained using a HgCdTe 1024x1024 infrared array
(CamIV) mounted on the Coud\'e focus of the 1.6 m telescope at the Observat\'orio do Pico
dos Dias/LNA (Brazil).  The stars were observed on 2001 June 09--12, and again
on 2004 April 30 and 2004 July 02--04. Spectra of variable stars were also taken on 2002
November 04, and 2003 June 28. 

For each object, we combined up to 35 individual exposures varying between 10s and 60s
to reach typically S/N=80 in the continuum. We used a 600 gr/mm grating and a 1-arcsec
slit width to  obtain R$\simeq$7000 and 0.64\,{\AA}/pixel of dispersion, covering the
spectral range of 10470--11000\,{\AA}. The sky+thermal background subtraction was
accomplished by dithering the star along the slit in 7 positions. We median-combined
these 7 frames to obtain a median sky image, which was then subtracted from each original frame. Each
sky-subtracted image was then divided by a normalized flat-field image. The following data reduction,
extraction of the spectrum and analysis was made by using usual IRAF routines. The wavelengths were
calibrated using an internal ThAr emission lamp, achieving a typical rms uncertainty of 0.08 {\AA}
($\sim$~2.2~km\,s$^{-1}$). The telluric features were removed from the program stars' spectra by
dividing them by the spectrum of a fast-rotating hot star. The photospheric Pa\,$\gamma$ and
\ion{He}{i}\,10912\,{\AA} absorptions were previously removed  from the fast-rotating star by
interpolating the adjacent continuum with a low-order Legendre polynomial. The extracted spectra were
corrected to the heliocentric frame and normalized to the continuum intensity using a low-order
Legendre polynomial (errors in normalization are typically 1\%). No flux calibration or de-reddening was done.

\section{The atlas\label{atlas}}

The atlas consists of 10 Figures, containing from two to six spectra of hot
luminous stars each, thus 40 peculiar objects. The
spectral evolution of selected objects during 2001--2004 is shown in Figs.
\ref{var1} to \ref{var3}. An additional Figure (Fig. \ref{early}) contains the spectra of normal early-type stars, to be
used as a reference for the characteristic photospheric features. Each Figure shows the air
wavelength in {\AA} versus the normalized intensity, plus an arbitrary offset, except where
indicated otherwise. Some figures were
re-sampled to show only the region that has strong spectral lines for a given group of stars. The
peculiar objects were grouped based on their spectroscopic characteristics, and the basic data of
these stars are presented in Table \ref{tabledata}. 

\begin{figure}
\resizebox{\hsize}{!}{\includegraphics{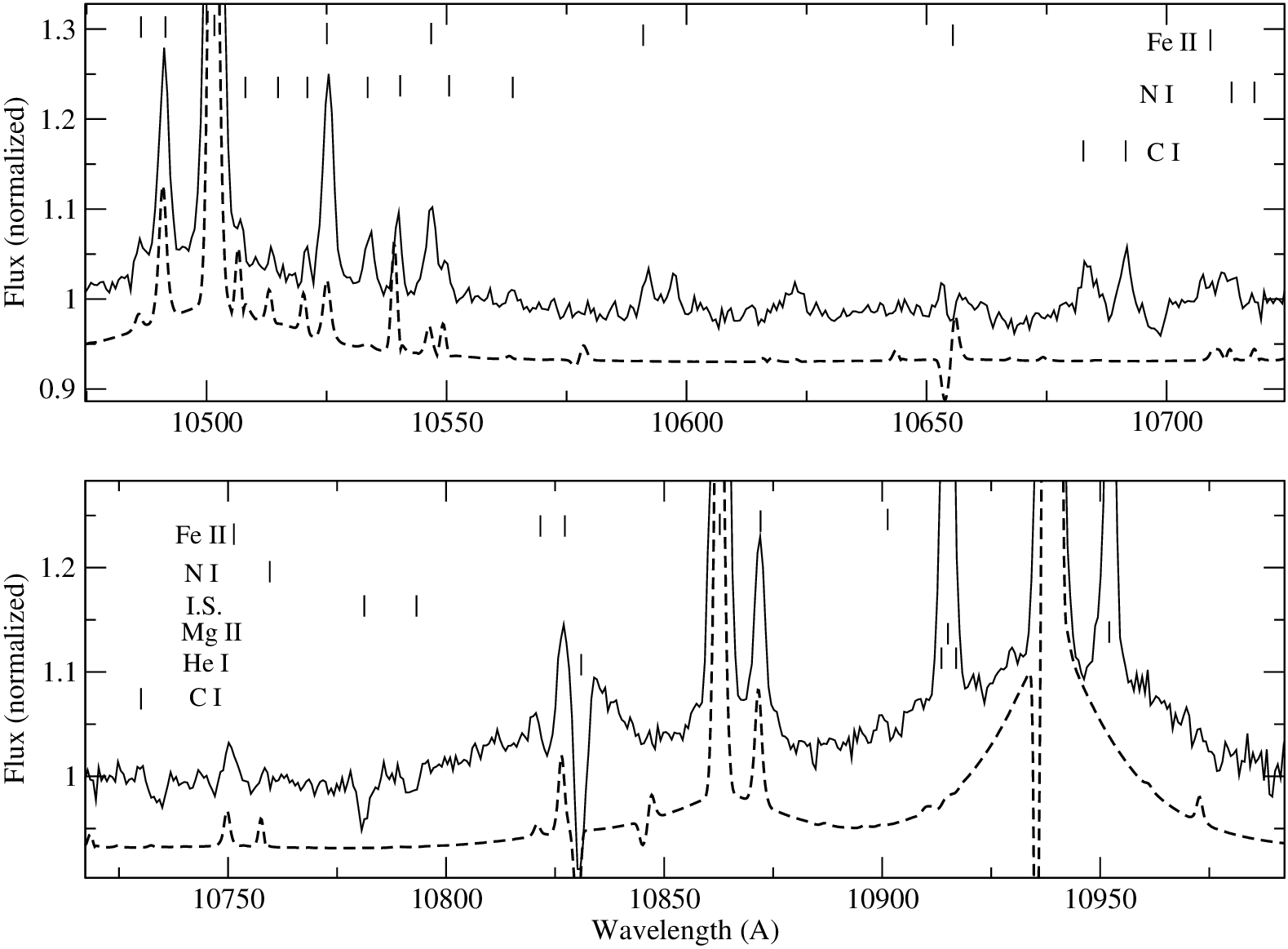}}
\caption{Identification of the features present in the 2004~April spectrum of the LBV 
AG~Carinae (full line). Thanks to the high resolution and high S/N provided by
the infrared spectrograph (CamIV) we could identify a number of weak features present in the spectrum. The dashed line
shows a non-LTE
radiative transfer model for AG~Car (J. H. Groh et al. 2006, in preparation) which was used to
confirm the line identifications. Extended, strong electron-scattering wings are noticeable in
\ion{He}{i}\,10830\,{\AA} and Pa\,$\gamma$, and are present in several other LBVs (see Figs.~\ref{lbv1}, 
~\ref{lbv2}, and \ref{lbv3}).}
\label{lineidenta}
\end{figure}

Many stars have been extensively studied in other spectral regions, whilst some of them
have very scarce recent bibliography. Compared to the previous published data in this
spectral region, the higher resolution and higher S/N of our spectra allowed us to identify many
faint spectral lines. They were identified by comparing the value of the oscillator strength of the
transition with the observed line intensities, and then checking for the presence of lines
of the same multiplet or ionization stage in other wavelength regions (e.\,g. optical).
Almost all the identifications were confirmed by comparing the observations with synthetic
spectra (J. H. Groh et al. 2006, in preparation), using the radiative transfer code CMFGEN
\citep{hm98}. However, there are still
unidentified spectral features remaining, which will be discussed in Sect.
\ref{discussion}. In Table \ref{tablelines} we summarize the detected features in the spectra of
the program stars; the  air wavelengths were taken from the Atomic Line List
v2.04\footnote{http://www.pa.uky.edu/~peter/atomic}. The line identifications are
overploted on the spectrum of the LBV AG Carinae taken in 2004 April (Fig.
\ref{lineidenta}).

\begin{table} 
\caption{Observed line transitions in the program stars.} 
\label{tablelines} 
\centering
\begin{tabular}{c c c}
\hline\hline
Ion &  $\lambda_{air}$ ({\AA}) & Identification\\
\hline
\ion{Fe}{ii}   & 10485.940 &  $\mathrm{e^6D_{5/2}-^6D^0_{5/2}}$     \\
\ion{Fe}{ii}   & 10490.942  & $\mathrm{z^4F_{7/2}-b^4G_{7/2}}$       \\
\ion{Fe}{ii}   & 10499.297  & $\mathrm{z^4F_{7/2}-b^4G_{5/2}}$       \\
\ion{Fe}{ii}   & 10501.499  & $\mathrm{z^4F_{7/2}-b^4G_{9/2}}$       \\
\ion{N}{i}     & 10507.000  & $\mathrm{^4P^0_3-^4D_5}$    \\
\ion{N}{i}     & 10513.410  & $\mathrm{^4P^0_1-^4D_1}$    \\
\ion{N}{i}     & 10520.580  & $\mathrm{^4P^0_3-^4D_3}$    \\
\ion{Fe}{ii}   & 10525.122  & $\mathrm{w^2D_{3/2}-^4P_{5/2}}$	      \\
\ion{N}{i}     & 10533.760  & $\mathrm{^4P^0_3-^4D_1}$        \\
\ion{N}{i}     & 10539.574  & $\mathrm{^4P^0_5-^4D_7}$        \\
\ion{Fe}{ii}   & 10546.348  & $\mathrm{e^6D_{9/2}-y^6F^0_{11/2}}$   \\
\ion{N}{i}     & 10549.640  & $\mathrm{^4P^0_5-^4D_5}$       \\
\ion{N}{i}     & 10563.330  & $\mathrm{^4P^0_5-^4D_3}$       \\
\ion{Fe}{ii}   & 10591.170  & $\mathrm{e^6D_{5/2}-^6D^0_{3/2}}$       \\
\ion{Fe}{ii}   & 10655.649  & $\mathrm{d^2F_{7/2}-z^2F^0_{7/2}}$       \\
\ion{He}{i}    & 10667.65   & $\mathrm{3p\,^3P^0-6s\,^3S}$     \\
\ion{C}{i}     & 10683.080  & $\mathrm{^3P^0_1-^3D_2}$      \\
\ion{C}{i}     & 10685.340  & $\mathrm{^3P^0_0-^3D_1}$      \\
\ion{C}{i}     & 10691.240  & $\mathrm{^3P^0_2-^3D_3}$       \\
\ion{C}{i}     & 10707.32  & $\mathrm{^3P^0_1-^3D_1}$       \\
\ion{Fe}{ii}   & 10709.730  & $\mathrm{e^6D_{7/2}-y^6F^0_{5/2}}$       \\
\ion{Fe}{ii}   & 10711.060  & $\mathrm{e^6D_{3/2}-^6D^0_{5/2}}$       \\
\ion{N}{i}     & 10713.548  & $\mathrm{^4P^0_3-^4P_5}$       \\
\ion{N}{i}     & 10717.950  & $\mathrm{^4P^0_3-^4P_3}$       \\
\ion{C}{i}     & 10729.53  & $\mathrm{^3P^0_2-^3D_2}$       \\
\ion{Fe}{ii}   & 10749.781  & $\mathrm{e^6D_{7/2}-y^6F^0_{7/2}}$      \\
\ion{N}{i}     & 10757.887  & $\mathrm{^4P^0_5-^4P_5}$      \\
\ion{Fe}{ii}   & 10820.88   & $\mathrm{e^6D_{3/2}-^6D^0_{3/2}}$       \\
\ion{Fe}{ii}   & 10826.483  & $\mathrm{e^6D_{7/2}-y^6F^0_{9/2}}$       \\
\ion{Fe}{ii}   & 10829.550  & $\mathrm{z^4D_{3/2}-b^4G_{5/2}}$	      \\
\ion{He}{i}    & 10830.34   & $\mathrm{2s\,^3S-2p\,^3P^0}$ \\
\ion{Fe}{ii}   & 10862.644  & $\mathrm{z^4F_{5/2}-b^4G_{7/2}}$       \\ 
\ion{Fe}{ii}   & 10871.601  & $\mathrm{z^4F_{5/2}-b^4G_{5/2}}$        \\
\ion{Fe}{ii}   & 10899.370  & $\mathrm{e^6D_{3/2}-y^6F^0_{1/2}}$       \\
\ion{He}{i}    & 10912.99   & $\mathrm{3d\,^3D-6f\,^3F^0}$                \\
\ion{Mg}{ii}   & 10914.280  & $\mathrm{^2D_5-^2P^0_3}$       \\
\ion{He}{i}    & 10917.062  & $\mathrm{3d\,^1D-6f\,^1F^0}$      \\
Pa\,$\gamma$ &  10938.095  & 6-3            \\    
\ion{Mg}{ii}   & 10951.770  & $\mathrm{^2D_3-^2P^0_1}$     \\
\hline
\end{tabular}
\end{table}

The atlas is organized as follows: in subsection \ref{be} we present the spectra of the
Be stars, in subsect. \ref{lbvs} we show the spectra of the LBVs and LBV candidates, in
subsect.
\ref{obsuperg} we group the OB supergiants, and in subsect. \ref{wr} the
spectra of the Wolf-Rayet stars. In each subsection we briefly introduce
the general characteristics of the class, and discuss the relevant aspects and
details of the observed features for each star, comparing them with another objects
of our atlas when appropriated.

\subsection{The Be stars \label{be}}

The 13 classical Be stars of our sample are shown in decreasing order of spectral type 
in Figs. \ref{be1}, \ref{be2}, and \ref{be3}. Most stars
in this class show \ion{Fe}{ii} lines, \ion{He}{i} 10830 {\AA} and Pa\,$\gamma$  in emission, often
double-peaked. The sole exception is HD~120991, which does not show \ion{Fe}{ii} lines. Some of the Be
stars also show emissions in \ion{C}{i} and \ion{Mg}{ii} lines, that have the same morphology as the
\ion{Fe}{ii} emission.

{\bf X Persei }(Fig. \ref{be3}) is a Be/X-ray binary in our sample, such as HD~110432
\citep{clark01}. X Persei is the optical 
counterpart of the pulsar 4U 0352+30 \citep{clark01}. The system has a low eccentricity
(e=0.11) and a long orbital period of 250 days \citep{dm01}. Its
optical spectrum is characterized by both H$\alpha$ and \ion{He}{i} 6678 {\AA} showing significant
variability \citep{clark01}. In the near-infrared, its spectrum shows double-peaked emission in \ion{Fe}{ii} 10500
{\AA} (-100 and 195 km\,s$^{-1}$) , while \ion{He}{i} 10830 {\AA} presents a weak
double-peaked profile (-98 and 13 km\,s$^{-1}$), such as Pa\,$\gamma$ (-100 and -5 km\,s$^{-1}$). \ion{Fe}{ii} 10862 {\AA}
is present too, but it is very weak. \ion{Mg}{ii} lines at 10915 and 10952 {\AA} are
narrow. 

{\bf $\delta$ Scorpii}  (Fig. \ref{be3}) has been claimed to be a classic Be\,star based on
its optical and infrared spectroscopy history, recently undergoing a H$\alpha$
outburst \citep{fabregat02}. However, for many years it was
considered a typical B-type star. Previous observations did  not show any emission
lines \citep{heasley83,grigsby92}. $\delta$ Scorpii is in a highly eccentric binary
system, with an orbital period of 10.6 years \citep{banerjee01}.  Since the appearance
of emission lines coincided with the periastron passage, 
binarity could be an alternative hypothesis to explain the Be phenomenon in this object \citep{fabregat00,mir01}.
In our near-infrared spectrum $\delta$ Scorpii shows double-peaked \ion{Fe}{ii} 10500 {\AA} and 10862 {\AA}
(-65 and +130 km\,s$^{-1}$ and -45 and +80 km\,s$^{-1}$, respectively).\ion{C}{i}
 10683, 10685 and 10691 {\AA} (blend), \ion{C}{i} 10707 {\AA}, and \ion{C}{i} 10729 {\AA}  are
present in emission. The \ion{He}{i} 10830 {\AA}  line is strong, while Pa\,$\gamma$ is weak. The 
\ion{Mg}{ii} lines are broad and may be double-peaked. 

{\bf HD~110432 }(Fig. \ref{be3}) is a Be star that is located just beyond the southern part of the
Coalsack at a distance of 300 $\pm$ 50 pc \citep{rachford01}. It is one of
the brightest stars lying  behind interstellar material and it has narrow
absorption resonance lines in  the UV spectrum \citep{codina84,prinja87}. 
The optical data displays line-profile variations from night to night.
\citet{torrejon01} classified HD~110432 as a low luminosity Be/X-ray binary.
They found a pulsation period of 14 ks in X ray emission, suggesting that the
system is a Be\,+\,White\,Dwarf candidate. Its near-infrared spectrum looks like
that of $\delta$ Scorpii, although we have found a symmetrically double-peaked
profile in \ion{He}{i} 10830 {\AA} and in Pa\,$\gamma$, both with peaks at -80 and 80 km\,s$^{-1}$. 
As in $\delta$ Centauri, a weak feature was found at 10541 {\AA} and we identified
it as \ion{N}{i} 10539.6 {\AA}. 

{\bf $\chi$ Ophiuchi} (Fig. \ref{be2}) displays fast variability in the Balmer series
lines \citep{doazan70}. \citet{borisova92} observed variations in the \ion{He}{i} optical lines
with a period of 0.913\,day, and suggested that $\chi$ Oph is a non-radial pulsator.
Its near-infrared spectrum is dominated by narrow emissions lines of \ion{Fe}{ii}, \ion{Mg}{ii},
\ion{C}{i} and Pa\,$\gamma$. The \ion{C}{i} lines are more prominent than in the other Be stars of our
sample. The \ion{He}{i} 10830 {\AA} emission is weak. 

{\bf HD~120991} (Fig. \ref{be2}) is a pole-on Be star that has shown strong and
variable emission features since it was first studied by \citet{fleming1890}.
The optical H lines show strong profile variations and the \ion{Fe}{ii} lines show weak V/R
variability \citep{han95,han96}. Its IUE spectrum was
first studied by \citet{dachs84}, who claimed that most of the resonance lines arise
in the interstellar medium. The high excitation resonance lines of \ion{N}{v}, \ion{C}{iv}, and \ion{Si}{iv}
do not seem to be formed in a high-velocity expanding envelope \citep{hubeny86}. 
Hence, he concluded that the IUE spectrum of HD~120991 is associated with the stellar
photosphere. In the near-infrared spectrum this star presents few narrow emission lines (\ion{He}{i}
10830 {\AA}, Pa\,$\gamma$ and \ion{Mg}{ii}) that have a single-peak profile. 

{\bf $\upsilon$ Cygni} (Fig. \ref{be2}) is a classical Be star first detected by
\citet{fleming1890}. This object shows strong  H$\alpha$ emission \citep{neiner05}, variable on
different timescales. Outbursts are quite common in this object, such as the one detected by the
Hipparcos satellite. In our atlas its spectrum resembles that of $\delta$ Centauri, except that the
\ion{C}{i} lines are stronger, \ion{Mg}{ii} lines are weaker, and Pa\,$\gamma$ is symmetrically
double-peaked (-50 and +55 km\,s$^{-1}$). The \ion{Fe}{ii} 10500 {\AA} and 10862 {\AA} lines are
double-peaked as well (-30 and +115 km\,s$^{-1}$ and -36 and +77 km\,s$^{-1}$, respectively).

{\bf 66 Ophiuchus} (Fig. \ref{be1}) is a well-observed Be star. \citet{peters82} analyzed the first IUE
observations and detected high velocity absorption components (from -250 to -850 km\,s$^{-1}$) in the
resonance lines of \ion{C}{iv}, \ion{Si}{iv}, and \ion{Si}{iii}. Based on the UV spectrum, strong variations were detected in
its stellar-wind lines  \citep{barker85}, vanishing and appearing in intervals of several months
\citep{grady87}. 66 Oph has a long history of optical variability, showing H$\alpha$  emission with
irregular variations. Cyclic V/R variability was also detected in H$\alpha$ during the period 1982--1993.
In our near infrared spectrum 66 Oph shows weak \ion{Fe}{ii} , \ion{C}{i}  and \ion{He}{i} emission
lines. Pa\,$\gamma$ is almost symmetrically double-peaked (violet and red peaks at -110 and +90
km\,s$^{-1}$, respectively), reaching up to 0.4 of the continuum intensity (i.e., F$_{peak}$/F$_c$=1.4).
Pa\,$\gamma$ also presents an absorption profile on the blue side that can be P~Cygni absorption.

{\bf $\eta$ Centauri }(Fig. \ref{be1}) is a bright and fast-rotating star
(v\,sin\,i=310~km\,s$^{-1}$ ), and is a member of the Scorpio-Centaurus OB association
\citep{degeus89}. It rotates at 0.65 of its critical velocity \citep{hutchings79} and the estimated
mass-loss rate from Si IV doublet profiles is log~$\dot{M}=-9.55~$M$_{\sun}$\,yr$^{-1}$ \citep{snow82}.
\citet{stefl95} found a nearly sinusoidal variation in the radial velocity of
\ion{Si}{iii} 4552 {\AA} with a period of
0.6424~day. In the near-infrared spectrum $\eta$ Cen presents weak lines of \ion{C}{i} and \ion{Mg}{ii}. 
\ion{Fe}{ii} 10500 {\AA} is symmetrically double-peaked  (-175 and 260 km~s$^{-1}$), while \ion{Fe}{ii}
 10862 {\AA} and \ion{Fe}{ii} 10872 {\AA} are blended. \ion{He}{i} 10830 {\AA} displays a shell feature
(central intensity of F/F$_c$=0.4), indicating that this object is seen nearly at the equator.  Pa\,$\gamma$ shows a
complex structure, with three emission peaks.

{\bf $\delta$ Centauri} (Fig. \ref{be2}) is a member of the Sco-Cen OB association, 
and \citet{fleming1890} again was the first to detect its emission-line
features. \citet{han95} found V/R variability in H$\alpha$ and \ion{Fe}{ii} lines, which he
proposed to be due to global disk oscillations. \citet{clark98} identified $\delta$ Cen
as a radio source, which may be variable. In the near-infrared region it shows a rich
variety of \ion{Fe}{ii} lines. \ion{Fe}{ii} 10490  {\AA} and \ion{Fe}{ii} 10525 {\AA} are weak (F$_{peak}$/F$_c$=1.05),
while \ion{Fe}{ii} 10500 {\AA} appears double peaked (-60 and +80
km\,s$^{-1}$). The \ion{Fe}{ii} 10862 {\AA} profile is double-peaked too (-88 and + 30
km\,s$^{-1}$). We found weak emission lines at 10541 {\AA} and 10550 {\AA} which we
identified as \ion{N}{i} 10539.6 {\AA} and 10549.6 {\AA}, respectively. The blend of \ion{C}{i} lines
is weak (F$_{peak}$/F$_c$=1.05), peaking at 10684 {\AA} and 10692
{\AA}. The \ion{He}{i} 10830 {\AA} line shows a broad  profile, reaching up to F$_{peak}$/F$_c$=1.15.
Pa\,$\gamma$ is strong (F$_{peak}$/F$_c$=1.25) and the peak is
slightly redshifted (+25 km\,s$^{-1}$). The \ion{Mg}{ii} lines are present as well. 

{\bf $\alpha$ Arae} (Fig. \ref{be1}) does not show V/R variability in the optical spectrum, presenting
only a constant and moderated Balmer emission \citep{mennickent91,han95}. In our spectrum it presents
double-peaked \ion{Fe}{ii} emission both in 10500 {\AA} (-105 and +142 km\,s$^{-1}$) and 10862 {\AA} (-60 and
+80 km\,s$^{-1}$). As almost all Be stars of our atlas, it displays a broad emission at 10686 {\AA} that
we identified as a blend of \ion{C}{i} lines (10683 {\AA}, 10685 {\AA} and 10691 {\AA}). An unidentified
absorption feature was found at 10906 {\AA}. $\alpha$ Arae also shows weak emission of \ion{Mg}{ii} 10915
{\AA} and 10952 {\AA}, and very weak (F$_{peak}$/F$_c$=1.05) emission of \ion{He}{i} 10830 {\AA}. Pa\,$\gamma$
is strong and double-peaked (-109 and +85 km\,s$^{-1}$). 

{\bf $\lambda$ Pavonis }(Fig. \ref{be3}) is a intermediate-rotating Be star
(v~sin~i=140~km\,s$^{-1}$ ).  This object revealed highly-variable H$\beta$ emission episodes
during 1984--1987  \citep{mennickent91}. Our monitoring at Observat\'orio Pico dos Dias
have shown many episodes of fading and reappearance of H$\alpha$ during the last 15\,years (A.
Damineli 2007, in preparation).
Its optical spectrum shows strong, double-peaked H emission \citep{sahade88}.
\citet{chen89} analyzed its IUE spectra and found that the profiles of the UV lines are
rotationally broadened into two groups, with rotational velocities of 170 and 210 km\,s$^{-1}$. They
inferred two different regions in the extended stellar atmosphere, one that is rotating and
one expanding, carrying the wind out at a maximum velocity of 155 km\,s$^{-1}$.
Only absorption lines are found in the IUE spectra of $\lambda$ Pav, ranging from neutral
species (e.g., \ion{C}{i} and \ion{N}{i}) to highly ionized ions (\ion{C}{iv} and \ion{Si}{iv} ). 
This object was detected as a
hard X-ray source by the Einstein Observatory and might have a neutron star as a companion.
The only line present in its near-infrared spectrum is \ion{He}{i} 10830 {\AA}, with
F$_{peak}$/F$_c$=1.45, and symmetrically  double-peaked (-130 and
+130 km\,s$^{-1}$).

{\bf HD~142983 }(Fig. \ref{be1}) is a V/R variable star with an extended atmosphere. It shows cyclical
variations with a period of about 10 years. In the UV spectrum it does not show any emission lines, and
the optical H and He lines are broadened by rotation \citep{mclau61,dc76}. \citet{floquet96} showed
multiple-frequency patterns in the  line-profile variations in HD~142983. They suggested that non-radial
pulsations and enhanced mass-loss rates are linked in Be stars. The near-infrared  spectrum of HD~142983
shows \ion{Fe}{ii} 10500 {\AA} symmetrically double-peaked (-135 and +127 km\,s$^{-1}$) and \ion{Fe}{ii} 10862 {\AA}
double-peaked too, but non-symmetrically (-150 and +95 km\,s$^{-1}$). \ion{C}{i} and \ion{Mg}{ii} lines are present,
but very weakly. \ion{He}{i} 10830 {\AA} shows a shell-like absorption profile (F/F$_c$=0.3), indicating that
HD~142983 is seen nearly equator-on. The same unidentified feature found in $\alpha$ Arae at 10906
{\AA} was found here as well. Pa\,$\gamma$ shows a shell profile, consisting of two emission shoulders separated
by a strong absorption feature, centered at -40 km\,s$^{-1}$. 

{\bf $\beta$ Monocerotis A} (Fig. \ref{be3}) is the brightest member of a complex 
triple system where the three components are Be stars of similar spectral
type \citep{mar94}. $\beta$ Mon A shows V/R variability with a
period of 12.5 years (McLaughlin 1961). \citet{mara91} reported  
profile variation of \ion{Mg}{ii} 4481 {\AA}, showing episodic emissions. Its
near-infrared spectrum resembles HD~142983, except for some unidentified
features in the former. It shows \ion{Fe}{ii} 10500 {\AA} double-peaked (-48 and +190 km\,s$^{-1}$) as well as
\ion{Fe}{ii} 10862 {\AA} (-80 and +135 km\,s$^{-1}$). \ion{Fe}{ii} 10525 {\AA} is also clearly present.
We detected a broad emission around 10689 {\AA} which we identified as a
blend of \ion{C}{i} (10683,10685,10691 {\AA}). We also detected narrow emissions at
10714, 10723 {\AA}, which we related to a double-peaked emission of  \ion{N}{i} 10718 {\AA} (-70
and +150 km\,s$^{-1}$, respectively). We found a broad emission at 10740 {\AA} and very
narrow absorptions at 10757 {\AA} and 10770 {\AA} that remain unidentified. \ion{He}{i} 10830
{\AA} displays a shell absorption (F/F$_c$=0.65), indicating that this object is seen
nearly equator-on. Pa\,$\gamma$ is double-peaked (-70 and +110 km\,s$^{-1}$), presenting a strong 
absorption centered at 30 km\,s$^{-1}$. The line intensity drops from F/F$_c$=2.0
in the Doppler-shifted peaks to F/F$_c$=1.4 in the dip.

\subsection{O, B, and A Iape Stars - or LBVs, LBV candidates and dormant LBVs} \label{lbvs}

The spectra of OBA Iape stars are presented in Figures \ref{lbv1}, \ref{lbv2}, \ref{lbv3} and
\ref{bpec1}. We grouped peculiar objects that share common morphological properties in the same figure,
although this is almost impossible in some cases. The stars in this group are all related to the LBV
phase; some of them are known as bona-fide LBVs (such as AG~Car and HR~Car), some are LBV candidates
(W\,243 in Westerlund~1, MWC~314) and some are believed to be in a dormant/post-LBV phase (such as He
3-519). For a recent review on the properties of these objects see 
\citet{hd94}, \citet{vg01}, and \citet{clark05}.

{\bf HD~168625 } (Fig. \ref{lbv1}) is classified as a marginally-dormant LBV based on its last 25-year
lightcurve \citep{sterken99}, although a typical LBV nebula was discovered around it
\citep{hut94,nota96,robberto98,pasquali02,ohara03}. Its optical spectrum was recently published by
\citet{chentsov03}, while \citet{hanson96} showed the K-band spectrum. Our spectrum
shows weak emission lines of \ion{Fe}{ii}, \ion{Mg}{ii} and NI. \ion{He}{i} 10830 {\AA} shows a deep P Cyg absorption with
a very weak emission, while Pa\,$\gamma$ is seen in absorption. As in HR Car, we also detected the 10780
{\AA} and the 10792 {\AA} absorption features. The comparison between the 2001 and 2004 datasets  do not show any evidence of variations in the spectral lines, confirming its current
dormant stage.

{\bf HD~168607 }(Fig. \ref{lbv1}) is only $62''$ away from HD~168625 in the sky, and
they are both believed to be members of the M 17 association, lying at 2.2 kpc
\citepalias{wf2000}. At this distance they would be separated by only
0.7 pc in projection, and regarding their similar blue spectra, it
has been proposed that they could be a pair of massive twin stars (see discussion
in \citetalias{wf2000}). The 25-year lightcurve of HD~168607 presented by \citet{sterken99}
shows the same qualitative behavior of other LBVs, such as AG Car,
but on a much longer timescale. In our atlas, the near-infrared spectrum of HD
168607 resembles that of HD~168625, although the former shows more developed
emission lines, suggesting a higher activity and a higher
mass-loss rate for HD~168607. This objected  has also shown spectroscopic variations in the last 3
years (Fig. \ref{var3}), with an increase in the absorption of \ion{He}{i} 10830 {\AA}
and an overall decrease in the intensity of the emission lines.

{\bf HD~160529 }(Fig. \ref{lbv1}) is a bright A-type hypergiant star. \citet{sterken91}, 
through Str\"omgren $uvby$ and near-infrared photometry,  found a change in the
spectral type from A9 to B8 between 1983 and 1991, and a brightness decrease of
0.5 mag in this period. Based on this behavior, they classified HD~160529 as an
LBV star, and analyzed its lightcurve variations as well. Its spectral
variability was analyzed by \citet{wolf74}, who found variations in the
Balmer lines profile and in the radial velocities of about 40 km\,s$^{-1}$.
More recently \citet{stahl03} reported a long-term spectroscopic monitoring 
and the recent lightcurve of HD~160529. They found little variation in the
spectral type around A2Ia from 1991 to 2002, mainly using \ion{He}{i} 5876 {\AA} as a diagnostic line. In our
atlas, the near-infrared spectrum of HD160529 is very similar to the spectrum
of HR Car, except that the former shows stronger emission in \ion{He}{i} 10830 {\AA}, with a P Cyg
profile. Moreover, Pa\,$\gamma$ is seen  in absorption. From 2001 to 2004 we detected a
radial velocity variation of 55 km\,s$^{-1}$. We also noticed an overall increase in the
strength of the absorption lines, with a new absorption feature appearing at 10581
{\AA}, and a drastic decrease in the \ion{Fe}{ii} emission. However, \ion{He}{i} 10830 {\AA} did
not change its strength (Fig. \ref{var3}).

{\bf HR Carinae }(Fig. \ref{lbv1}) is one of the rare bona-fide LBV stars in the Galaxy. Its variability
was first reported by Hertzsprung \citep{hoffleit40}, and since then several lightcurves have been published
\citep[see][]{vg97}. The high values of the luminosity and mass-loss rate of HR Car were first noted by
\citet{viotti71}, and the 1989--1990 spectral variability with the associated changes in the lightcurve
were analyzed by \citet{hut91}.  Their optical spectra show evidence of an expanding atmosphere with
multiple shells. The spectrum of the bipolar nebula associated with HR Car shows evidence of nitrogen
enrichment, as does the star itself. A detailed study of the optical nebula around HR Car is given in
\citet{nota97} and \citet{weis97}, and the mid-infrared imaging of the nebula is presented by
\citet{voors97}. The most recent analysis of the wind of HR Car was performed by \citet{machado02}, based on
spectra collected in April 1999. They derived an effective temperature of 10,000 K, a luminosity of
500,000 L$_{\sun}$ and a mass-loss rate of $6.5\times10^{-5}$ M$_{\sun}$\,yr$^{-1}$. Previously,
\citet{vg01} derived an effective temperature of 8,000~K for HR Car in 1992, while \citet{nota97} found
20,000~K (May 1995) and 15,000~K (January 1996).  The near-infrared spectra of HR Car in June 2001 shows weak
\ion{Fe}{ii} emission (10500 {\AA}, 10525 {\AA} 10862 {\AA} and 10871{\AA}). Pa\,$\gamma$ shows an
inverse P Cygni profile, while \ion{He}{i} 10830 {\AA} is almost absent, with a very weak broad
emission. HR Car shows a rich absorption spectrum of \ion{N}{i}, \ion{C}{i} and \ion{Mg}{ii}. An
unidentified feature was found at 10780 {\AA} and another at 10792 {\AA} (see Sect.~\ref{discussion}).
The data taken in 2004 show much more variability than in 2001 (Fig. \ref{var3}). The inverse P
Cyg profile of Pa\,$\gamma$ changed to a pure strong emission, with the equivalent width rising from 2
to 14 {\AA}. The \ion{Fe}{ii} lines became stronger too, while the \ion{Mg}{ii} lines that were present in
absorption turned into emission. The line of \ion{He}{i} 10830 {\AA} presented an intrincated profile,
with a double-peak emission superimposed on the previous broad emission.

{\bf W~243 }(Fig. \ref{lbv1}) is a recently-discovered LBV candidate in the massive, young open
cluster Westerlund~1 \citep{cn04}. Those authors presented an optical
spectrum characterized by a double-peaked emission of H$\alpha$ and
single-peaked emission in the \ion{He}{i} lines. They also reported the strengthening
of H$\alpha$ during 2001-2003. The comparison with older spectra revealed that
the star evolved from B2-5Ia to A2I from 1981 to 2003, suggesting an LBV
classification. A K-band spectrum obtained by \citet{groh06} in July 2005 suggests further evolution
towards a hotter temperature. 
In this work we obtained near-infrared data in 2004 that confirm its LBV
nature. Its spectrum is similar to other cool LBVs, showing a strong Pa\,$\gamma$
emission that resembles HR Car as it was in 2004. Another remarkable feature of
cool LBVs, namely, the presence of \ion{Mg}{ii}, \ion{N}{i} and \ion{C}{i} lines in absorption, are also seen in W243.
However, \ion{He}{i} 10830 {\AA} emission is present and stronger than in other cool LBVs. The
P-Cygni absorption of this line is noticeable, extending up to 400 km\,s$^{-1}$, which is much higher
than the expected wind terminal velocity for A supergiants. Further monitoring of W~243 is required to
address the origin of this high-velocity absorption.

{\bf $\eta$ Carinae }(Fig. \ref{lbv2}) is one of the most observed stars. It underwent a giant
eruption in the 1840s that generated the bipolar-shaped {\it Homunculus} nebula \citep{gaviola50}, and
another outburst in the 1890s that probably generated another nebula, the ''little Homunculus"
\citep{ishibashi03}. The reader is referred to \citet{dh97}, who reviewed the system and
its environment. Since then, many works related to the nature of the central source in $\eta$~Car have been
published, mainly regarding the 5.54-year cycle and the binary scenario \citep{damineli97} or alternative
ones \citep{davidson97}. Its near-infrared spectrum shows prominent \ion{Fe}{ii} emission at 10500 {\AA}, 10525 {\AA},
and 10862 {\AA}, with a broad and a narrow component. Pa\,$\gamma$ also shows the same shape of the
\ion{Fe}{ii} lines. \ion{He}{i} 10830 {\AA} is stronger, with a P Cygni profile extending up to -660
km\,s$^{-1}$. There are other absorptions at -380, -195 and -45 km\,s$^{-1}$. The changes in this
spectral region are dramatic over the 5.54-year cycle (Fig. \ref{var2}): a few days before 2003.5 (minimum
of the spectroscopic event, see \citeauthor{sd04} \citeyear{sd04}) the narrow emission components from
\ion{Fe}{ii}, \ion{He}{i} and Pa\,$\gamma$ shrunk until they faded completely.  The behavior of \ion{He}{i}
10830 {\AA} is striking: the emission components faded almost completely, while the absorption component of
the P Cyg profile extended up to -1300 km\,s$^{-1}$.

{\bf He~3-519} (Fig. \ref{lbv2}) is a galactic prototype of the WN11h class. Its
optical and ultraviolet spectrum is nearly identical to the well-known LBV AG
Carinae at minimum (i.\,e., during the hot phase), as discussed by \citet{sc94}
 and \citetalias{wf2000}. Although He~3-519 does not have the typical
photometric or spectroscopic behavior of a LBV, it shows a circumstellar
nebula that relates it to that class. Therefore, it is believed that He 3-519
belongs to a rare class of post-LBV, pre-WN stars \citep{davidson93}. The
near-infrared spectrum of He~3-519 is similar to that of HD~316285, but the former
displays a stronger \ion{He}{i} 10830{\AA} emission and is almost free of \ion{Fe}{ii}
lines. The 2001-2004 data evolution shows significant strengthening of \ion{He}{i}
10830 {\AA}, but without changes in the velocity of the edge of the P Cyg
absorption (Fig. \ref{var2}), resembling the pattern of the variations of HD~316285.

{\bf HD~316285 }(Fig. \ref{lbv2}) is a luminous, evolved massive star as discussed in
\citet{hillier98}. Based on a detailed spectral analysis, those authors found evidence for nitrogen and
helium enhancement on the stellar surface, and a large mass-loss rate of $2.4\times10^{-4}$
M$_{\sun}$\,yr$^{-1}$ was derived. Although there is
no evidence of photometric variability, \citeauthor{hillier98} also noticed
spectroscopic variation, specially in the \ion{Fe}{ii} lines. The near-infrared
spectrum of this star resembles that of $\eta$ Car, except that the nebular and
high-excitation lines
seen  in the former are not present in HD~316285. Also, \ion{He}{i} 10830
{\AA} does not show the additional absorption components seen in the spectrum
of $\eta$ Car. Regarding the variability of HD~316285, our 3 datasets from 2001 to
2004 show significant strengthening of the emission lines (Fig. \ref{var2}),
although no variations are noticed in the velocity of the wind (as seen from the P-Cyg
absorption).

{\bf MWC 314 }(Fig. \ref{lbv3}) is a scarcely-studied emission line star. It was first
identified by \citet{merrill27}, while \citet{allen73} reported a late-type spectral energy distribution
for this object. It was recognized as a high-luminosity B[e] star by \citet{mir96},
who estimated its stellar parameters (log L/L$_{\sun}=6.2$, $T=30\,000$~K, M=80~M$_{\sun}$) and
derived a high reddening (A$_V$=5.7 mag). A comparison between B[e] supergiants and
LBVs led him to propose that MWC 314 is an LBV candidate. A high resolution optical
spectrum of this star is presented by \citet{mir98}; its optical spectrum is 
dominated by emissions of \ion{H}{i}, \ion{He}{i} and \ion{Fe}{ii}.
Its near-infrared spectrum
resembles AG Car and P Cygni. However, its spectroscopic evolution shows an
intriguing behavior in \ion{He}{i} 10830 {\AA}, with a variable high-velocity absorption
wing extending up to ~1500 km\,s$^{-1}$ (Fig. \ref{var3}). This behavior is striking because such 
a high velocity is not expected for this kind of object. In Fig.~5 of \citet{mir98} a high-velocity
absorption can be seen in \ion{He}{i} 5876 {\AA}, although errors in the continuum normalization may
affect the interpretation (A. Miroshnichenko, private communication). Moreover, the same qualitative
behavior appears in $\eta$~Car during periastron passage, although much more dramatically in the latter 
(see Fig.~\ref{var2}).
Additional multi-wavelenght observations
of MWC~314 using a finer temporal sampling are required to address the origin of this high-velocity
absorption component.
The equivalent width of the emission lines also changed by a factor of 1.5 from
2001 to 2004.

{\bf AG~Carinae} (Fig. \ref{lbv3}) is the ``Rosetta Stone" of the LBV class,
lying at 6 kpc from the Sun \citep{humphreys89}. AG Car and its environment have
been extensively studied in the last decades. The star is surrounded by a
bipolar nebula first reported by \citet{thackeray50}, which was ejected from the star
$\sim$ 10$^4$ years ago \citep{nota92,pacheco92,lamers01}. The photometric variability in the
last century was analyzed by \citet{vg97}, while the recent
spectroscopic variability was studied by \citet{leitherer94} and \citet{stahl01}. 
The morphology and kinematics of the ejecta were analyzed by \citet{nota95},
and the nebular abundances were determined by \citet{smith97}. \citet{shore96}
found a color excess of E(B-V)=0.65 for AG Car based on IUE
data. The optical spectra at minimum were analyzed by \citet{sc94},
who found an enhanced helium abundance at the surface of the star. Recently, \citet{ghd06}
found that AG Car is rotating very fast (up to 0.86 of its critical velocity), and
therefore it lies close to the so-called $\Gamma\Omega$ limit \citep{mm_omega00}. The AG Car
near-infrared spectrum shows prominent emission lines of \ion{Fe}{ii} and \ion{Mg}{ii}, and weak
emission of \ion{C}{i} and \ion{N}{i}. \ion{He}{i} 10667 and 10830 {\AA} show P Cygni profiles and
Pa\,$\gamma$ is seen in emission. The variability of the lines from 2001 (hot
phase) to 2004 (cool phase) is striking in this spectral region, specially in
\ion{He}{i} 10830 {\AA} (Fig. \ref{var3}). The line equivalent width dropped from 122
{\AA} in 2001 to 0.2 {\AA} in 2004. Also, in 2001 an additional
absorption component appeared in this line.

{\bf P Cygni }(Fig. \ref{lbv3}) is a famous member of the LBV class,
sharing its name to denote the characteristic shape of the spectral lines present
in those stars. P Cyg underwent a major eruption in the 1600s, lying in a
quiescent phase since then. This eruption (and perhaps a previous one)
gave rise to complex ejecta around the star \citep{skinner98,chesneau00,exter02}.
The atmospheric parameters derived by
\citet{najarro97} revealed a helium-enriched photosphere and confirmed the
high mass-loss rate and luminosity of the star. A conference regarding the
latest P Cyg progresses was organized in 2000; we refer the reader to the
proceedings of the event (ASP Conf. Proceedings 233) for a general review.
The near-infrared spectrum of P Cygni is very similar to AG Car at minimum.
It shows strong \ion{He}{i} 10830 {\AA} and Pa\,$\gamma$ emission, followed by more modest Fe
II and \ion{Mg}{ii} emission. From 2001 to 2003 the emission lines of P Cyg became
stronger, and since then they started to fade (Fig.
\ref{var3}).

{\bf HD~326823 }(Fig. \ref{bpec1}) is located in the direction of the Galactic Center at a
distance of 2 kpc \citep{mcgregor88}. \citet{sterken95} suggested that
HD~326823 could be a LBV in a quiescent state, based on their photometric study.
\citep{lopes92} presented the optical and the 1\,$\mu$m spectrum of this object. The
majority of the lines are double-peaked, probably indicating deviations from
spherical symmetry and suggesting the presence of a disk. The helium and nitrogen
lines are the strongest ones, while H$\alpha$ and H$\beta$ are weakly present. Hence,
they concluded that HD~326823 is an evolved star, entering the Wolf-Rayet phase.
\citet{borges01} present a high-resolution optical atlas of this star,
identifying the strongest features. The lines in the near-infrared spectrum show the same
double-peaked morphology as the optical ones. The strongest emission line seen is \ion{He}{i}
10830 {\AA}, followed by \ion{Fe}{ii} and \ion{Mg}{ii} lines. Pa\,$\gamma$ and \ion{C}{i} lines are barely seen.
Unfortunately, we observed HD~326823 only in the 2001 run, preventing us from
studying the variability of the star.

{\bf GG Carinae }(Fig. \ref{bpec1}) is suggested to be an eclipsing binary with a period
of 31 or 62 days \citep{hernandez81,gosset85}, although the
radial velocity curve is not well determined. The ultraviolet spectrum was
published by \citet{brandi87}, who recognize numerous \ion{Fe}{ii} and \ion{Ni}{ii}
features. Our near infrared data from 2001 shows \ion{Fe}{ii}, \ion{Mg}{ii}, \ion{C}{i} and \ion{He}{i}
10830 {\AA} in emission. Pa\,$\gamma$ is the strongest line in our spectrum, showing a
blueshifted feature at 10931.5 {\AA}, that we could not identify, which may be a
component of Pa\,$\gamma$ formed in a high-velocity shell.

{\bf HD~87643 }(Fig. \ref{bpec1}) is thought to be an evolved B[e] star \citep{oud98}
and appears to be related to the LBV class, although the
link is not clear yet. This object has an optical spectrum dominated by
emission lines of \ion{Fe}{ii} and P Cygni profiles in the Balmer series, together
with low excitation forbidden lines. The spectrum is produced by a fast polar
wind combined with a slow disk wind \citep{oud98}. The ultraviolet
and optical spectra of HD~87643 was previously discussed by \citet{pacheco82,pacheco85},
who reported a strong spectral line variability.
 HD~87643 has a bright reflection nebula, which was analyzed by \citet{surdej81} and
\citet{surdej83}. 
Its 2001 near infrared spectrum shows prominent \ion{Fe}{ii}, CI lines and Pa\,$\gamma$
emission, compatible with the presence of a  cold wind. \ion{He}{i} 10830 {\AA}
presents a P-Cygni profile, with a weak emission and a strong absorption that
goes up to -1750 km\,s$^{-1}$, which is probably formed in the fast polar wind.

\subsection{OB supergiants} \label{obsuperg}

In Figure \ref{obsuper} we present the OB supergiants of our sample, some of them which are related to
the LBV class. The stars are presented in decreasing order of spectral type. The spectra of the OB
supergiants typically show \ion{He}{i} 10830 {\AA} with a P Cyg profile and \ion{He}{i} 10667 {\AA}, He
I 10913 {\AA} and Pa\,$\gamma$ in absorption, with a variable veiling of their absorption depending on
how developed the wind is, culminating with HD~152408, that has only emission lines. The later subtypes
also present weak emission of \ion{Fe}{ii} 10862 {\AA}, \ion{Mg}{ii} 10915 {\AA} and \ion{Mg}{ii}
10952 {\AA}.

{\bf HD~152408} and {\bf HD~151804} (Fig. \ref{obsuper}) are closely-associated 
O supergiants in Sco OB1. The stellar wind of HD~152408 is stronger
despite the fact that the evolutionary
difference between both stars must be very small \citepalias[see][]{wf2000}. 
Both objects were analyzed by \citet{cb97} who derived
similar stellar parameters but a higher mass-loss rate and a
higher helium content in HD~152408 than in HD~151804, indicating the more evolved nature of the former. 
The morphology of their near-infrared spectra agrees with the analysis of \citet{cb97},
with HD~152408 having strong emission of \ion{He}{i} 10830 {\AA},
\ion{He}{i} 10913 {\AA} and Pa\,$\gamma$, while HD~151804 shows the same morphology as the
hotter B supergiants stars. An unidentified feature is seen at 10922.5 {\AA} in
emission in HD~152408, while in HD~151804 it is in absorption.

{\bf HD~154090} (Fig. \ref{obsuper}) is
thought to be a normal B supergiant star, with an abundance typical of
early-type stars \citep{kane81}. Indeed, the near-infrared spectrum that
we present is similar to other B supergiants. However, there are
significant variations between the 2001 and the 2004 datasets, especially in \ion{He}{i} 10830 {\AA}
 (Fig. \ref{var3}). In three years, the equivalent width of this
line reduced by a factor of two, with the P Cyg absorption almost disappearing.
These changes occurred together with radial velocity variations of $\sim$120
km\,s$^{-1}$, and minor changes in the absorption profiles of Pa\,$\gamma$ and \ion{He}{i} 10913 {\AA}.
This behavior encourages future monitoring of this star for better insight into its nature.

{\bf HD~169454 }(Fig. \ref{obsuper}) is a member of the Sct OB3 association, located
at 1.5 kpc from the Sun \citep{humphreys78}. However, the analysis of the
high-velocity interstellar gas in the direction of HD~169454 puts the star at
least at 3 kpc, which yields an absolute magnitude of M$_{\mathrm{V}}\sim$\,-9.2. As a
consequence, HD~169454 may be one of the brightest B-type stars in the Galaxy. Its
optical spectrum was analyzed by \citet{rivinius97}, who found a
variability pattern similar to that shown by HD~152236. We found that their near-infrared
spectrum and the changes from 2001 to 2004 are very similar as well (Fig.
\ref{var1}). 

{\bf HD~152236 }(Fig. \ref{obsuper}) is located in the Sco OB1 association, belonging
to the open cluster NGC 6231 \citep{vg84}. Photometric variability
in this object was reported by \citet{sterken77} and \citet{sterken97}, 
who suggested the relationship between the cyclic charactheristics of the lightcurve of HD~152236
and the LBV class. These authors also claim an increase in the apparent visual magnitude through the
past centuries, with the star being 2 magnitudes brighter in the 18$^{\mathrm {th}}$ century. \citet{burki82}
analyzed the UV variability, showing the presence of discrete absorption
components (DACs). \citet{rivinius97} investigated the changes in the
optical spectrum revealing an intense variability during 1990--91.  The
quantitative study of the spectrum by these authors also confirm that HD~152236
is a very luminous star (log\,$L/ L_{\sun} = 6.05$). Its near-infrared spectrum shows
\ion{He}{i} 10830 {\AA} with a P Cyg profile with an absorption extending up to $-450$ km\,s$^{-1}$ and \ion{He}{i}
10667 {\AA}, \ion{He}{i} 10913 {\AA} and Pa\,$\gamma$ in absorption. The comparison between
spectra taken in 2001 and 2004 shows radial
velocity changes of about 120 km\,s$^{-1}$ and in the profile of \ion{He}{i}~10830~{\AA}. 
Also, the emission was stronger and the absorption was shallower in 2001 when compared
to the 2004 dataset (Fig. \ref{var1}).

{\bf HD~80077 }(Fig. \ref{obsuper}) is a very luminous B star ($M_{\mathrm{Bol}}=-11$) with a 
comparatively low mass-loss rate of $5\times10^{-6} M_{\sun}\,yr^{-1}$ \citep{carpay89,carpay91}
among luminous B stars.  These authors
claim that the star could be an LBV, since it has a low surface gravity and a
high luminosity. Historical observations also suggest that the mass-loss rate was
higher some decades ago \citep{houk78}. The lightcurve presented by \citet{vg92}
shows that during 15 years the star had an apparent
visual magnitude of $m_{\mathrm{V}}=7.5$, with micro-variations of about 0.05 mag. It is still
not known why HD~80077 is such a stable star, while its LBV neighbors in the HR diagram are
unstable. We observed this star only in 2001, and its near-infrared
spectrum is very similar to other luminous B stars, such as HD~152236 and HD~169454. 
The only noticeable difference is the partial filling of the Pa\,$\gamma$
absorption and the presence of \ion{Mg}{ii} lines in emission. HD~80077 also displays
an unidentified absorption feature at 10700 {\AA}.

\subsection{Wolf-Rayet stars} \label{wr}

We have obtained spectra of 4 WN stars (Fig. \ref{wn}) and 3 WC stars  (Fig.
\ref{wc}). All of them are listed in the VIIth Catalog of Galactic Wolf-Rayet
stars \citep{vanderhucht01}, where their properties are extensively listed. 
Physical properties of WR stars are thoroughly reviewed in \citet{crowther07}.

The spectra of the WN stars, namely {\bf WR 6} (WN4), {\bf WR 136} (WN6(h)), {\bf WR 78 }
(WN7h) and {\bf WR 22} (WN7h+O9III-V), show a very strong P Cygni profile in \ion{He}{i} 10830
{\AA} blended with emission in Pa\,$\gamma$. The WNE stars WR 6 and WR 136 show
flat-topped line emission profiles, while the later WN stars of our sample
show a more round-topped line emission profile. In WR 22 it is also noticeable
the \ion{He}{i} 10830 {\AA} absorption from the O9 star.

Our sample of WC stars contains only the later subtypes ({\bf WR90}=WC7,
{\bf WR11}=WC8+O and {\bf WR103}=WC9d). This class of objects shows a broad, strong
emission of \ion{He}{i} 10830 {\AA} with a small P Cyg absorption. There is also a
blended emission from \ion{C}{iii} + \ion{C}{iv} centered at 10545 {\AA}. In the WC9d star
WR103 there is an emission of \ion{He}{i} 10913 {\AA} that indicates its cool nature
compared to other WC stars. In WR 11 ($\gamma^2$~Vel) 
the Pa\,$\gamma$ absorption from the secondary star (O7.5III-V) is also prominent.

\section{Discussion: unidentified absorption features} \label{discussion}

In the previous section we presented unambiguous identification of most emission lines present in
the spectral region covered by this work. The presence of \ion{C}{i} lines in Be
stars and LBVs is remarkable, as carbon is expected to be depleted in those objects, either
due to rotational mixing or by the ejection of the outer layers. The
near-infrared \ion{C}{i} lines potentially can be used to constrain the carbon abundance in these stars. 

\begin{figure} \resizebox{\hsize}{!}{\includegraphics{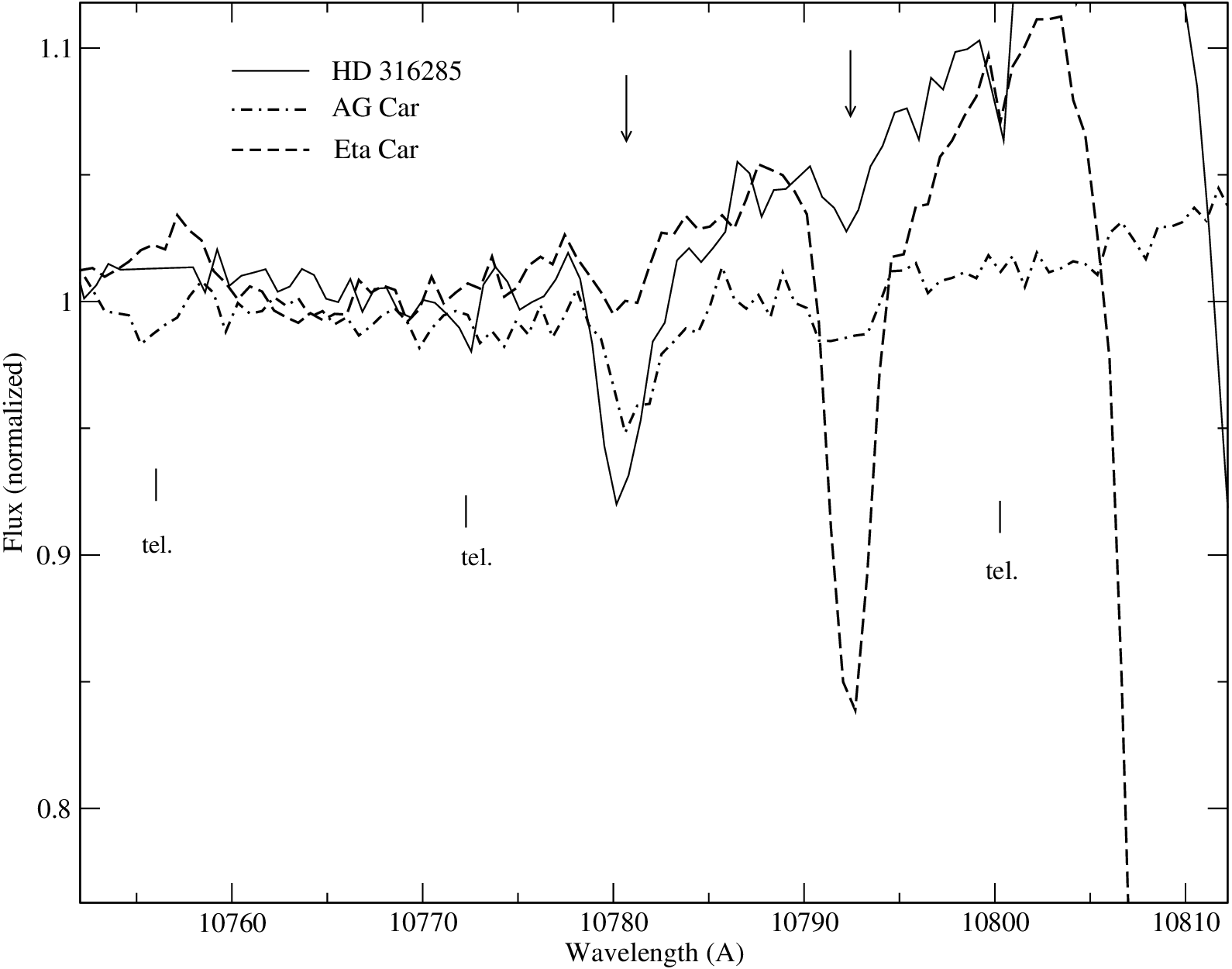}} \caption{Unidentified absorption lines
found in the LBVs HD~316285 (full), AG~Car (dot-dashed) and $\eta$~Car (dashed) at 10780 {\AA} and at 10792
{\AA} (arrows). The position of the strongest telluric lines are
also indicated, to show that they do not contaminate the absorption features. }
\label{10780a}
\end{figure}

\begin{figure}
\resizebox{\hsize}{!}{\includegraphics{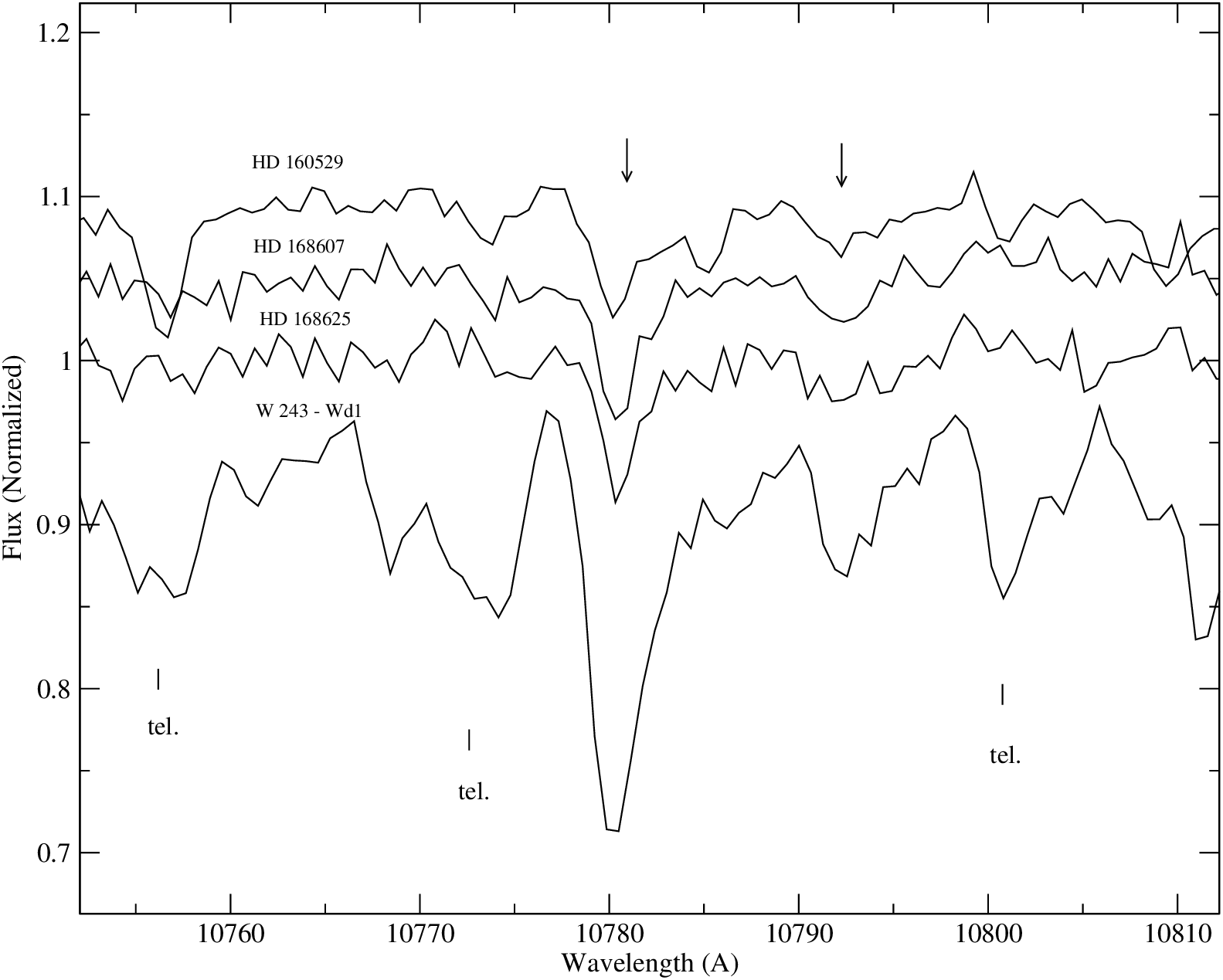}}
\caption{Same as Fig. \ref{10780a}, showing the presence of the same absorption lines in HD~160529,
HD~168607, HD~168625 and W243. For clarity, the spectra were shifted up by an arbitrary offset. }
\label{10780b}
\end{figure}

\begin{figure}
\resizebox{\hsize}{!}{\includegraphics{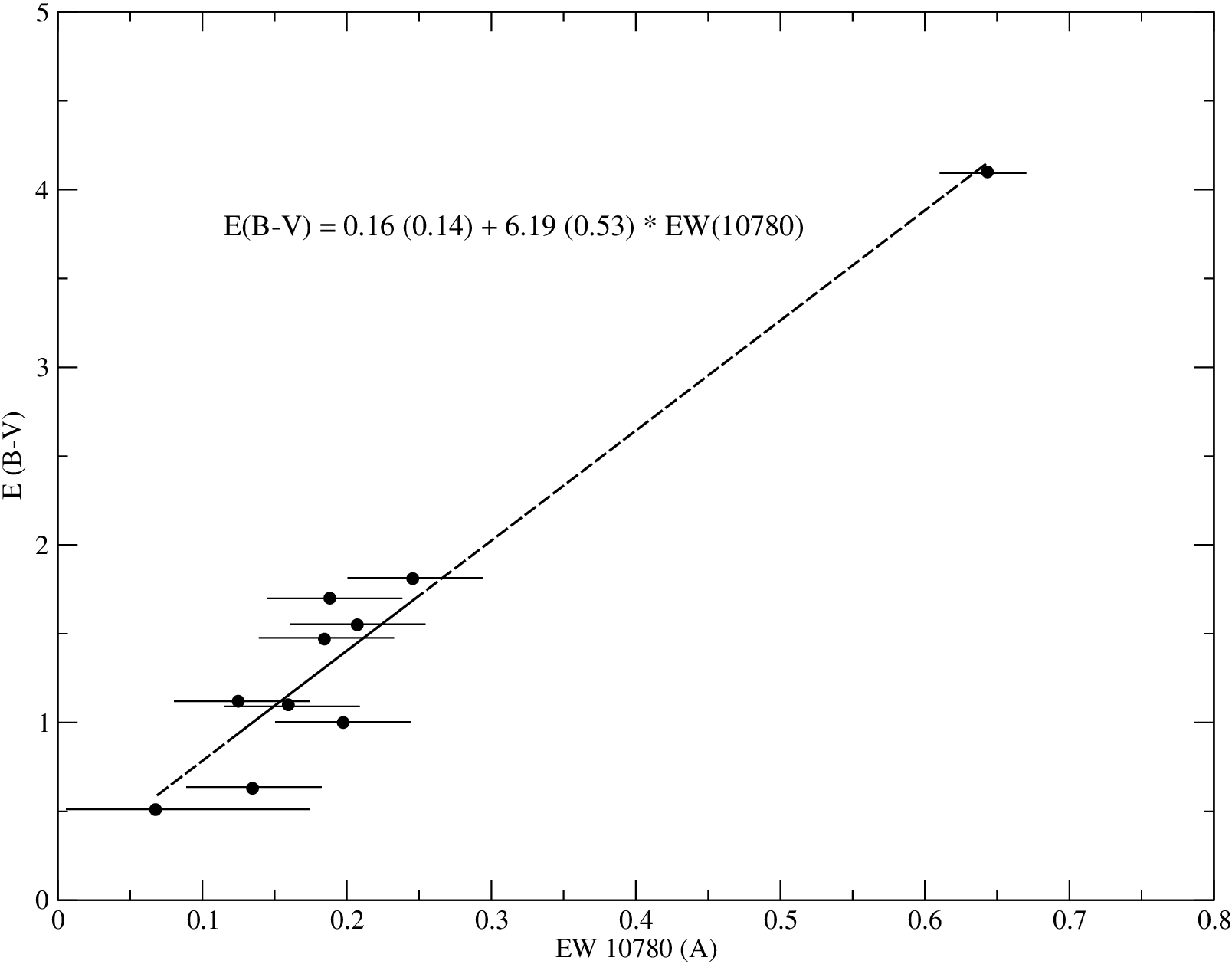}} \caption{Calibration of EW (10780)
versus E(B-V). The color excesses E(B-V) were taken from \citet{vg01},
except for W243 (M. Teodoro, private communication). A least-squares linear fit to the data is
also shown (see Eq.~\ref{10780_ebv_eq}).} \label{10780_ebv}
\end{figure}

\begin{figure*}
\centering
\scalebox{0.8}{\includegraphics[width=17cm]{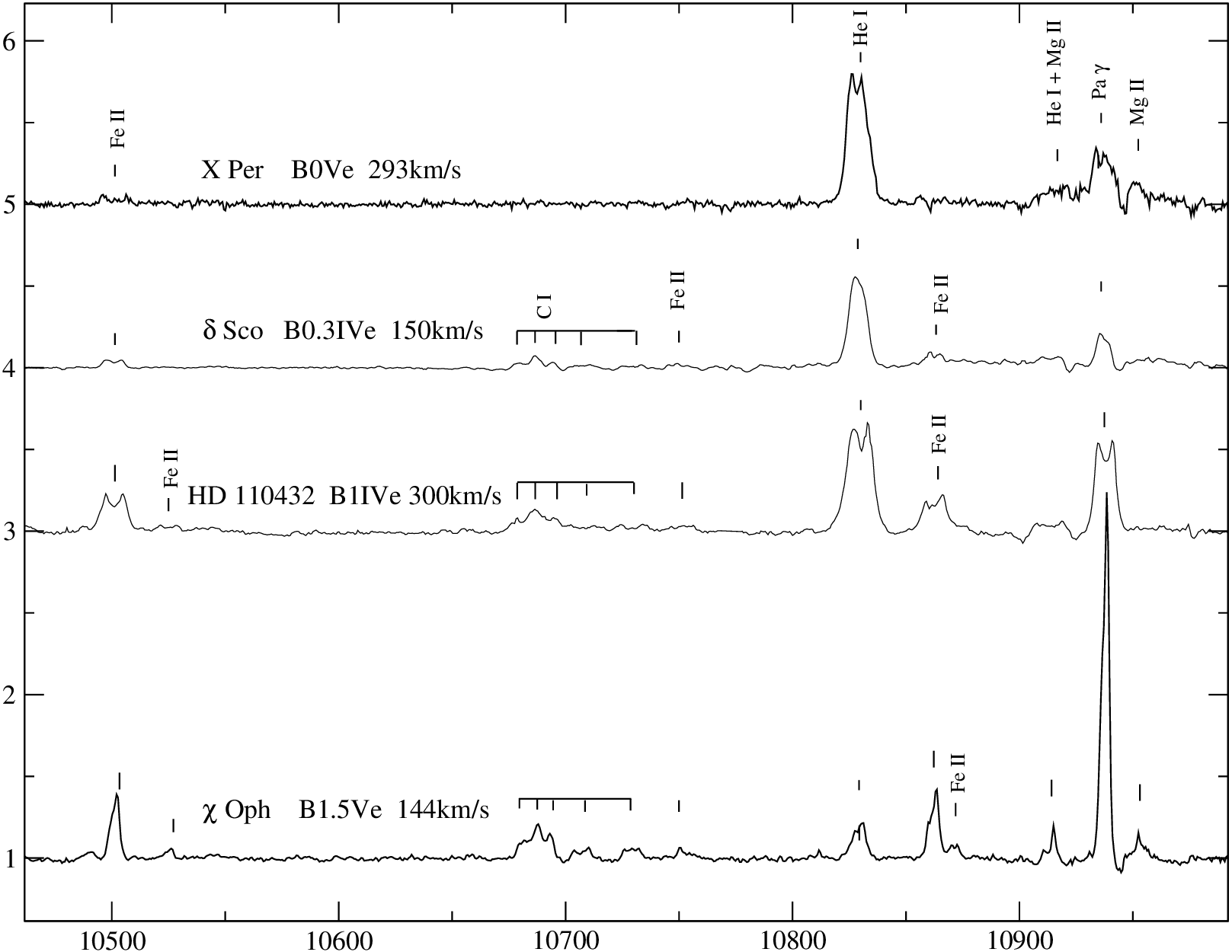}}
\caption{Spectra of the classical Be stars of our sample ordered by spectral type. Identification, spectral
type and  projected rotational velocity, respectively, are indicated for each object. From top to bottom, 
X Per, $\delta$~Sco, HD~110432,
and $\chi$~Oph. For clarity, the spectra were shifted by an arbitrary offset.}
\label{be1}
\end{figure*}

\begin{figure*}
\centering\scalebox{0.8}{\includegraphics[width=17cm]{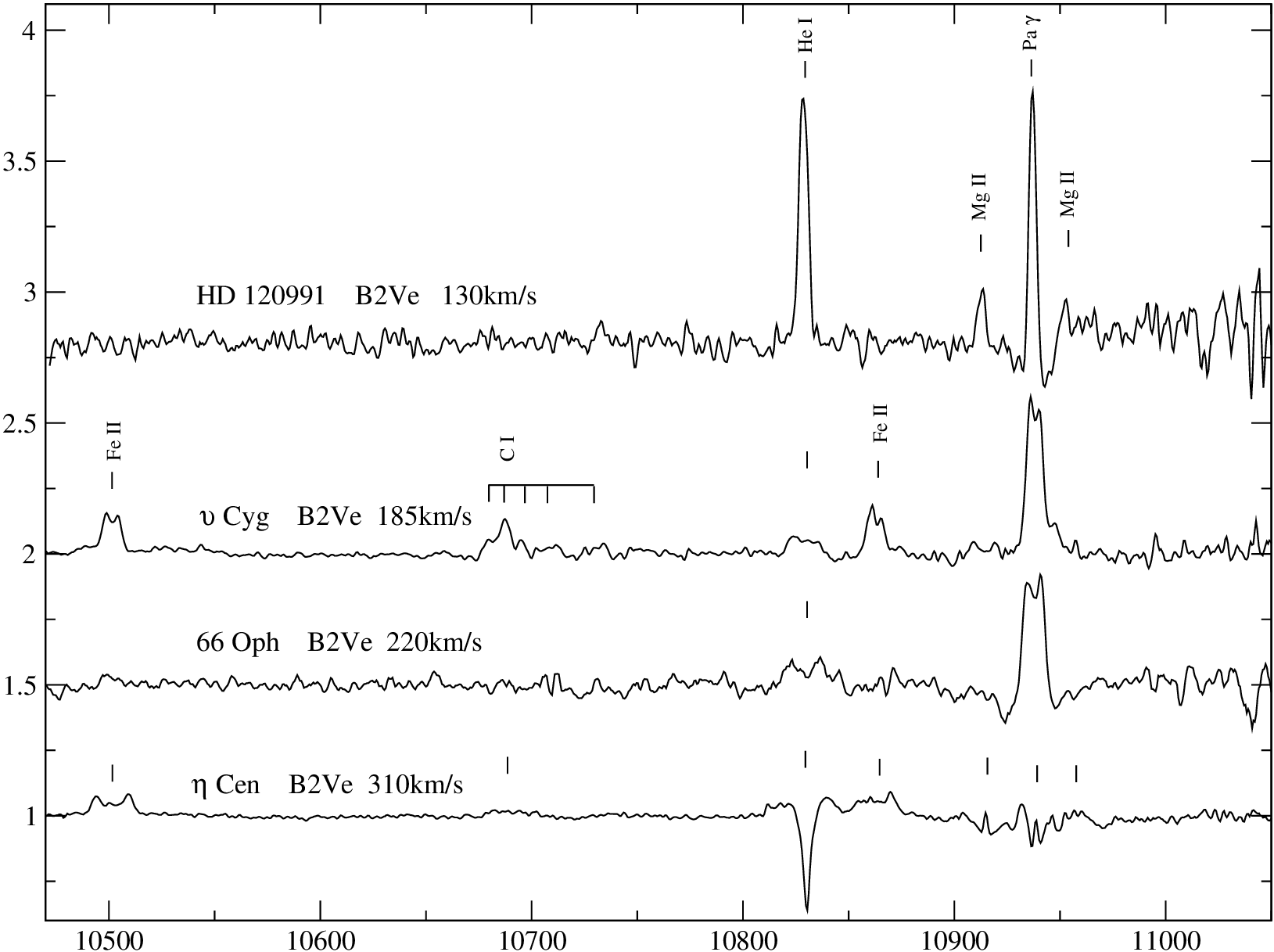}}
\caption{Same as Fig.~\ref{be1}, showing HD~120991, $\upsilon$ Cyg, 66~Oph, and $\eta$ Cen.  \label{be2}}
\end{figure*}

\begin{figure*}
\centering
\scalebox{0.8}{\includegraphics[width=17cm]{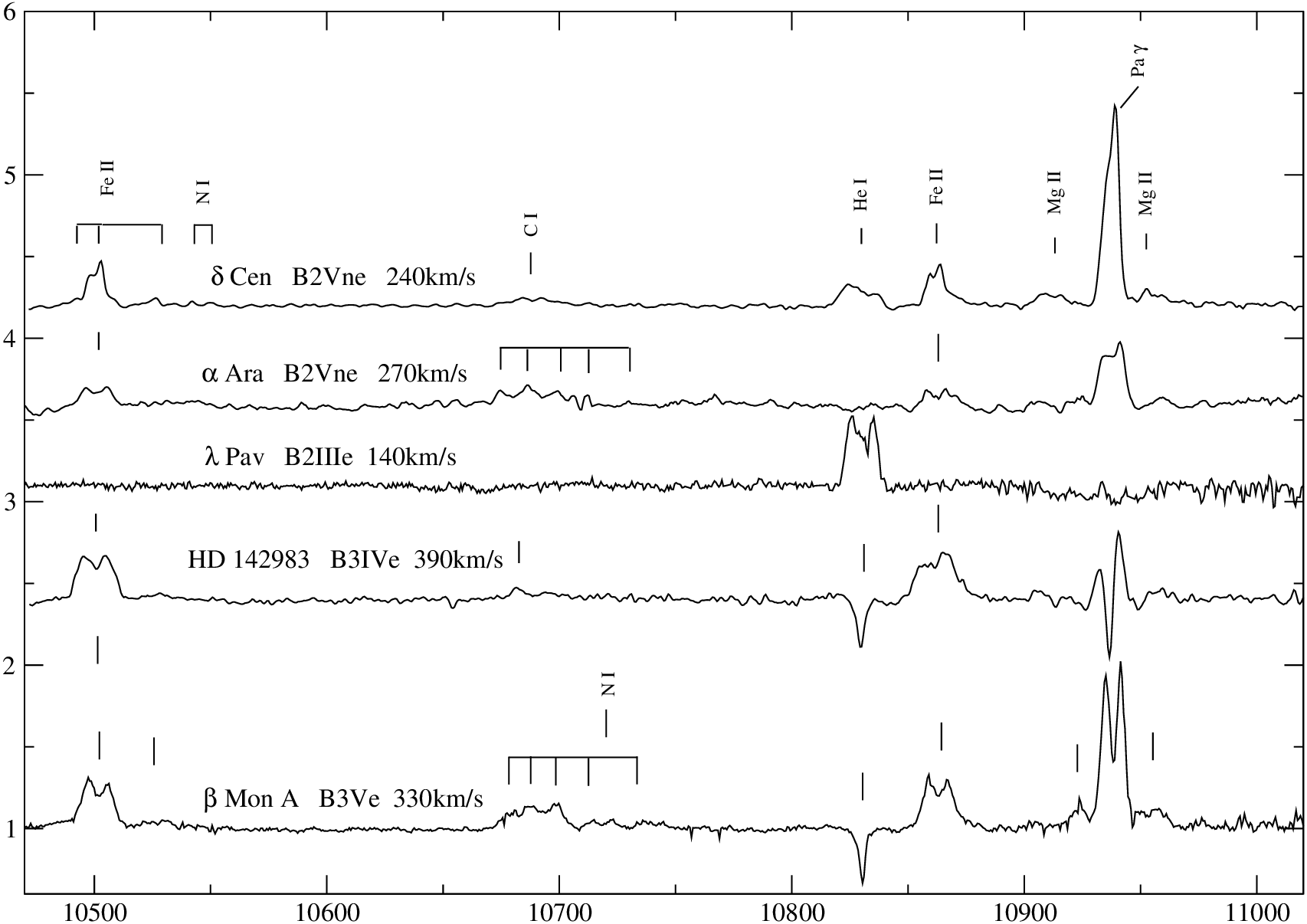}}
\caption{Same as Fig.~\ref{be1}, showing $\delta$~Cen, $\alpha$~Ara, $\lambda$~Pav, HD~142983, and $\beta$~Mon~A. 
 \label{be3}}
\end{figure*}

\begin{figure*}
\centering
\scalebox{0.8}{\includegraphics[width=17cm]{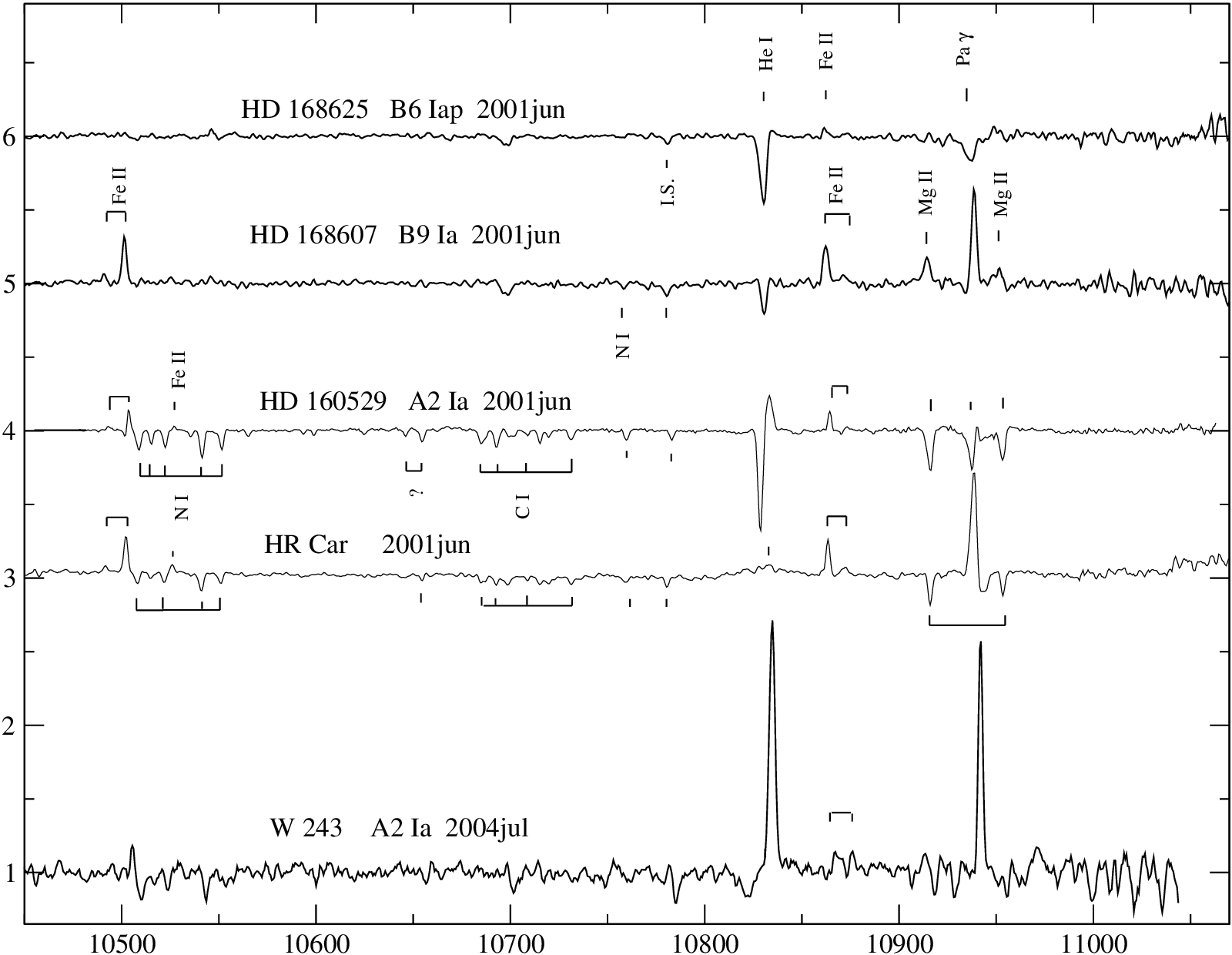}}
\caption{Spectra of LBVs and LBV candidates. From top to bottom, HD~168625, HD~168607,
HD~160529, HR~Car, and W243. Spectra of variable stars also give year and month of the observation. \label{lbv1}}
\end{figure*}

\begin{figure*}
\centering
\scalebox{0.8}{\includegraphics[width=17cm]{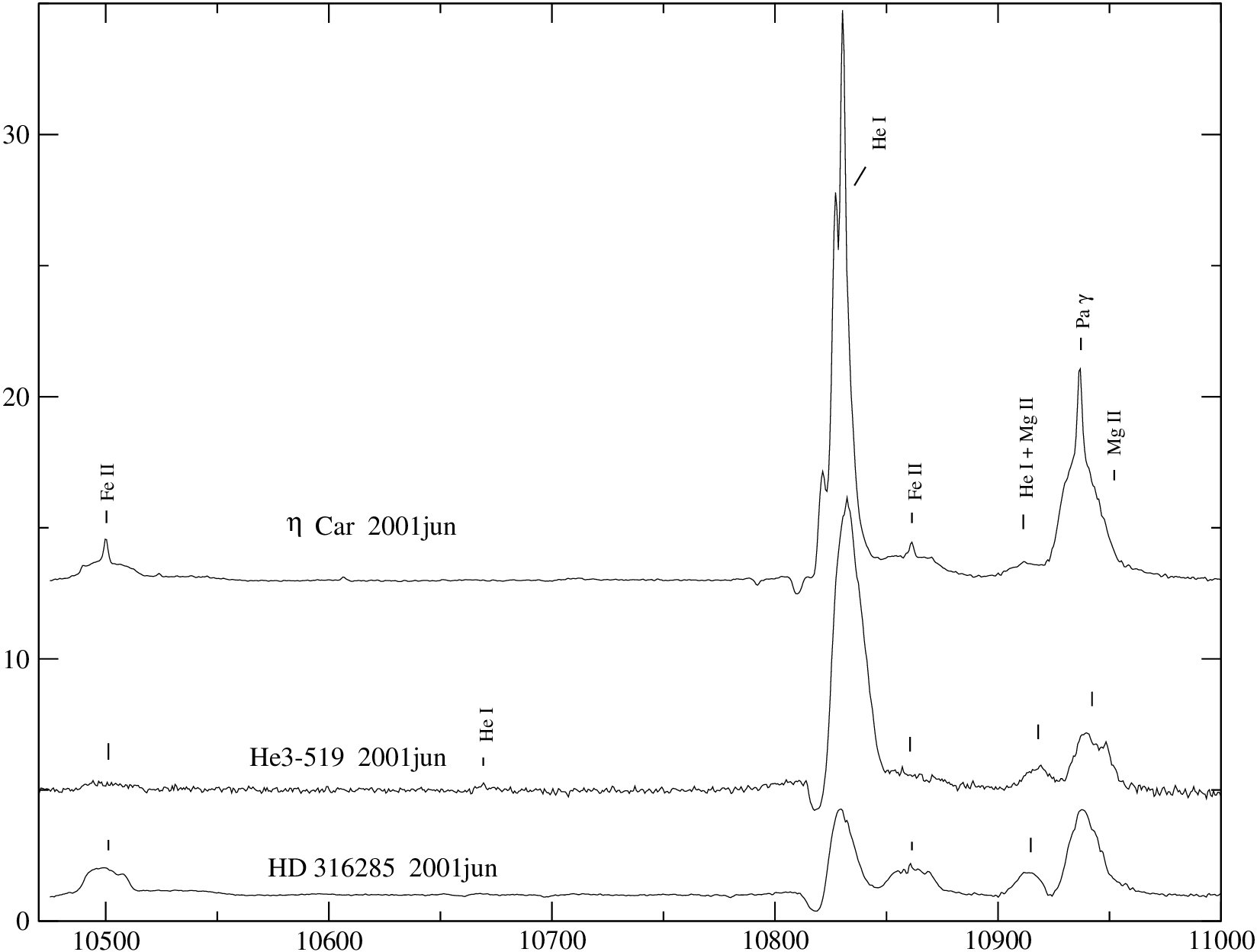}}
\caption{Same as Fig.~\ref{lbv1}, showing $\eta$~Car, He~3-519, and HD~316285. \label{lbv2}}
\end{figure*}

\begin{figure*}
\centering
\scalebox{0.8}{\includegraphics[width=17cm]{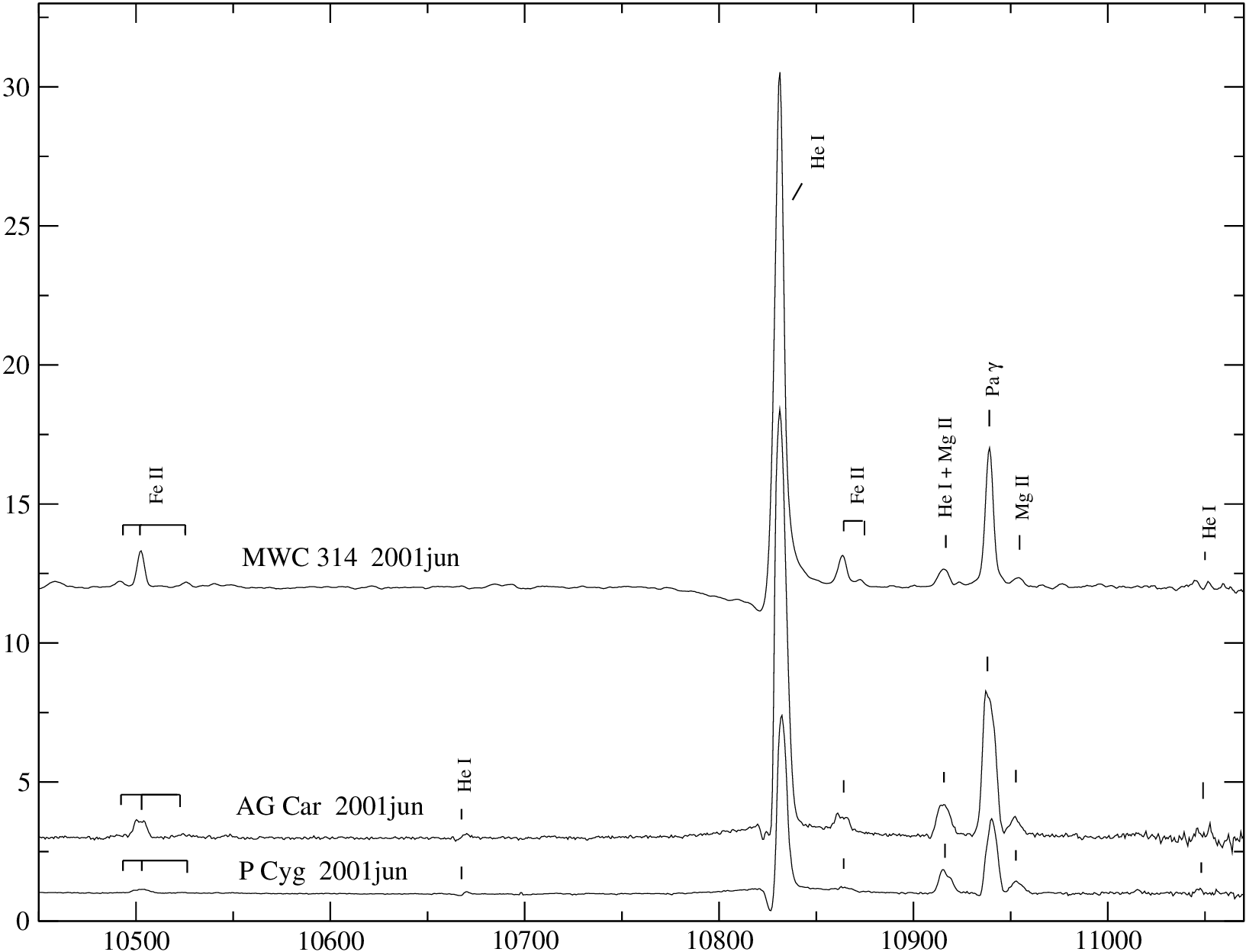}}
\caption{Same as Fig.~\ref{lbv1}, showing MWC~314, AG~Car, and P~Cyg. \label{lbv3}}
\end{figure*}

\begin{figure*}
\centering
\scalebox{0.8}{\includegraphics[width=17cm]{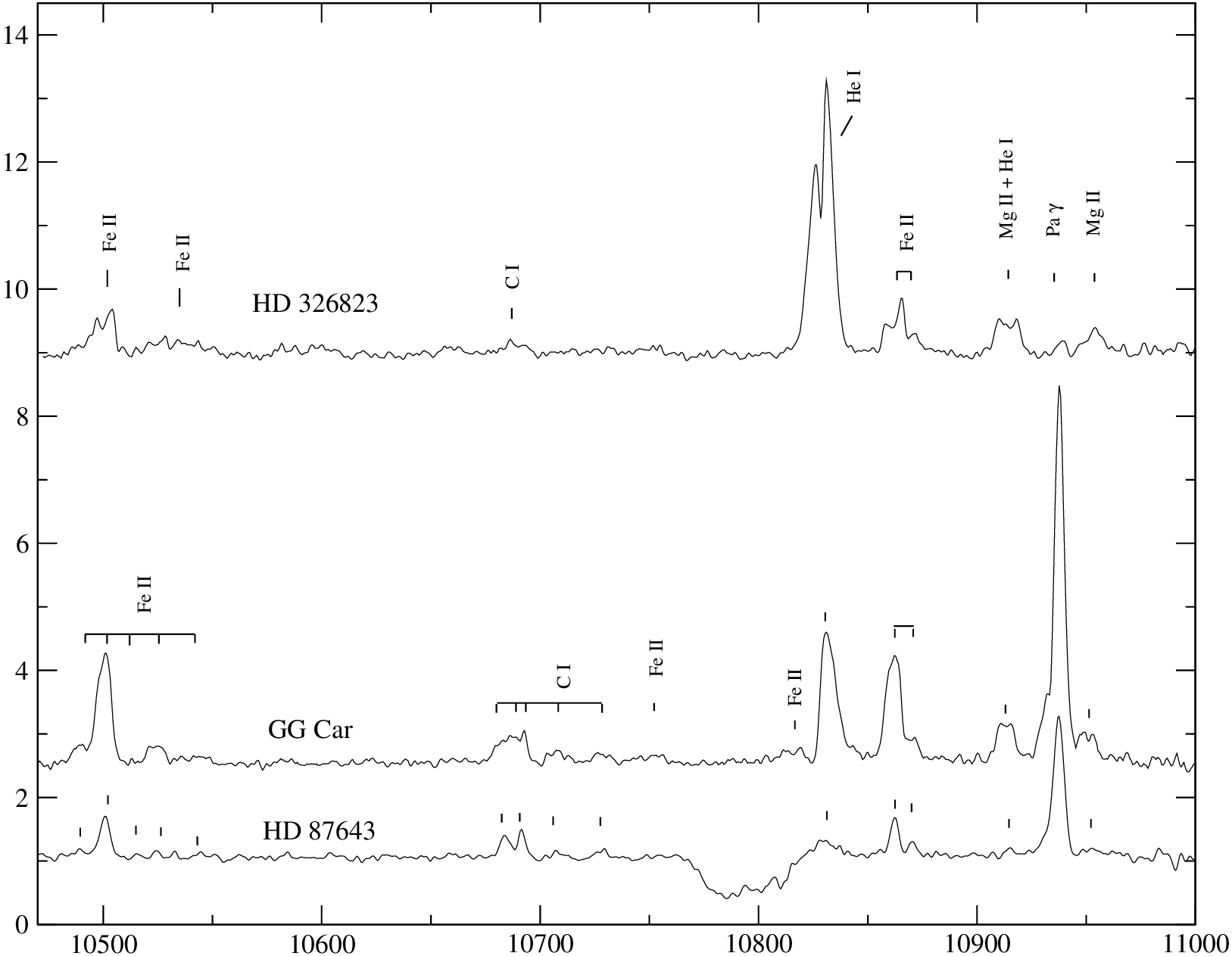}}
\caption{Same as Fig.~\ref{lbv1}, showing HD~326823, GG~Car, and HD~87643. \label{bpec1}}
\end{figure*}

\begin{figure*}
\centering
\scalebox{0.8}{\includegraphics[width=17cm]{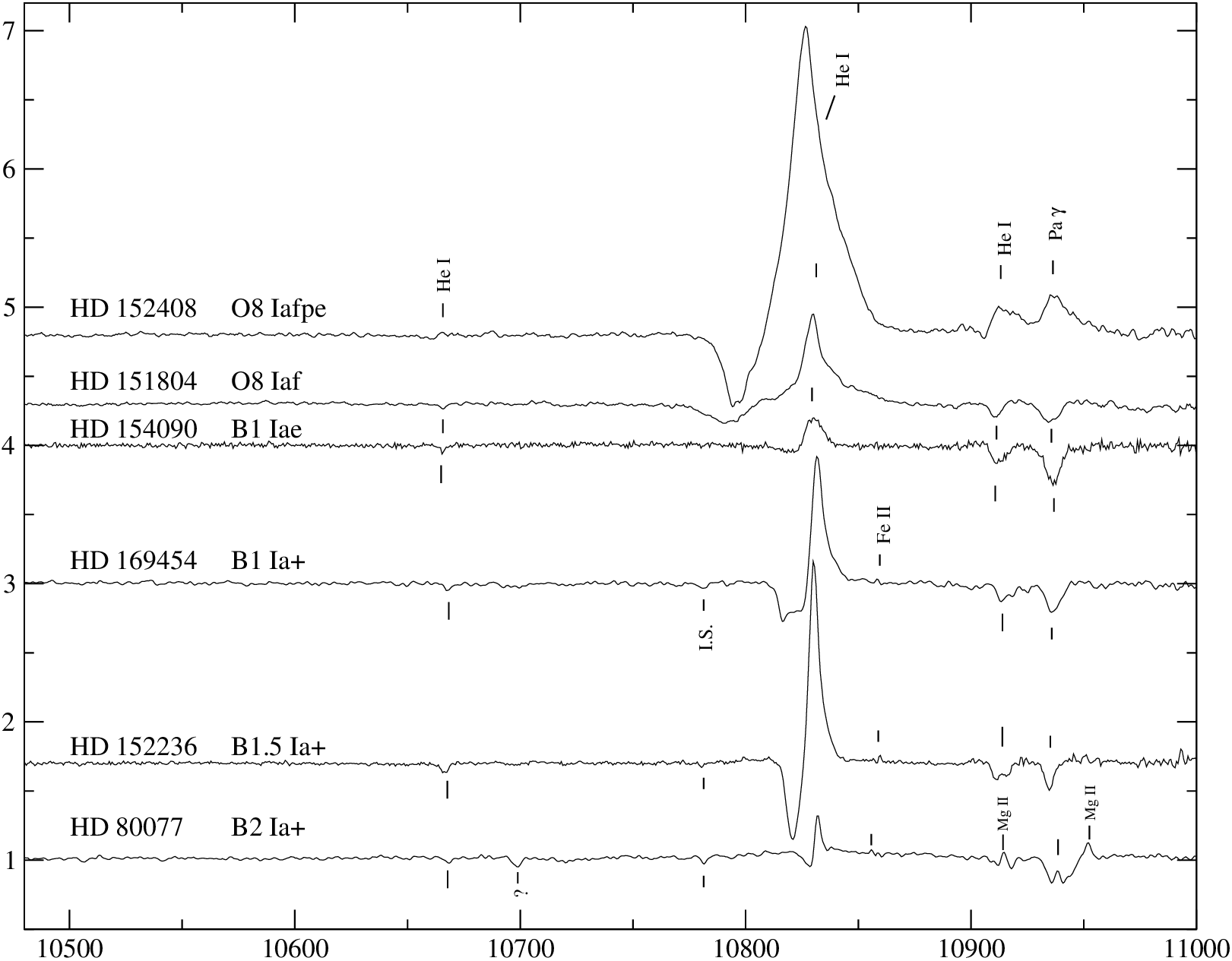}}
\caption{Spectra of the OB supergiants of our sample, showing from top to bottom HD~152408,
HD~151804, HD~154090, HD~169454, HD~152236, and HD~80077. The spectra were offset
by an arbitrary offset.
\label{obsuper}}
\end{figure*}

\begin{figure*}
\centering
\scalebox{0.8}{\includegraphics[width=17cm]{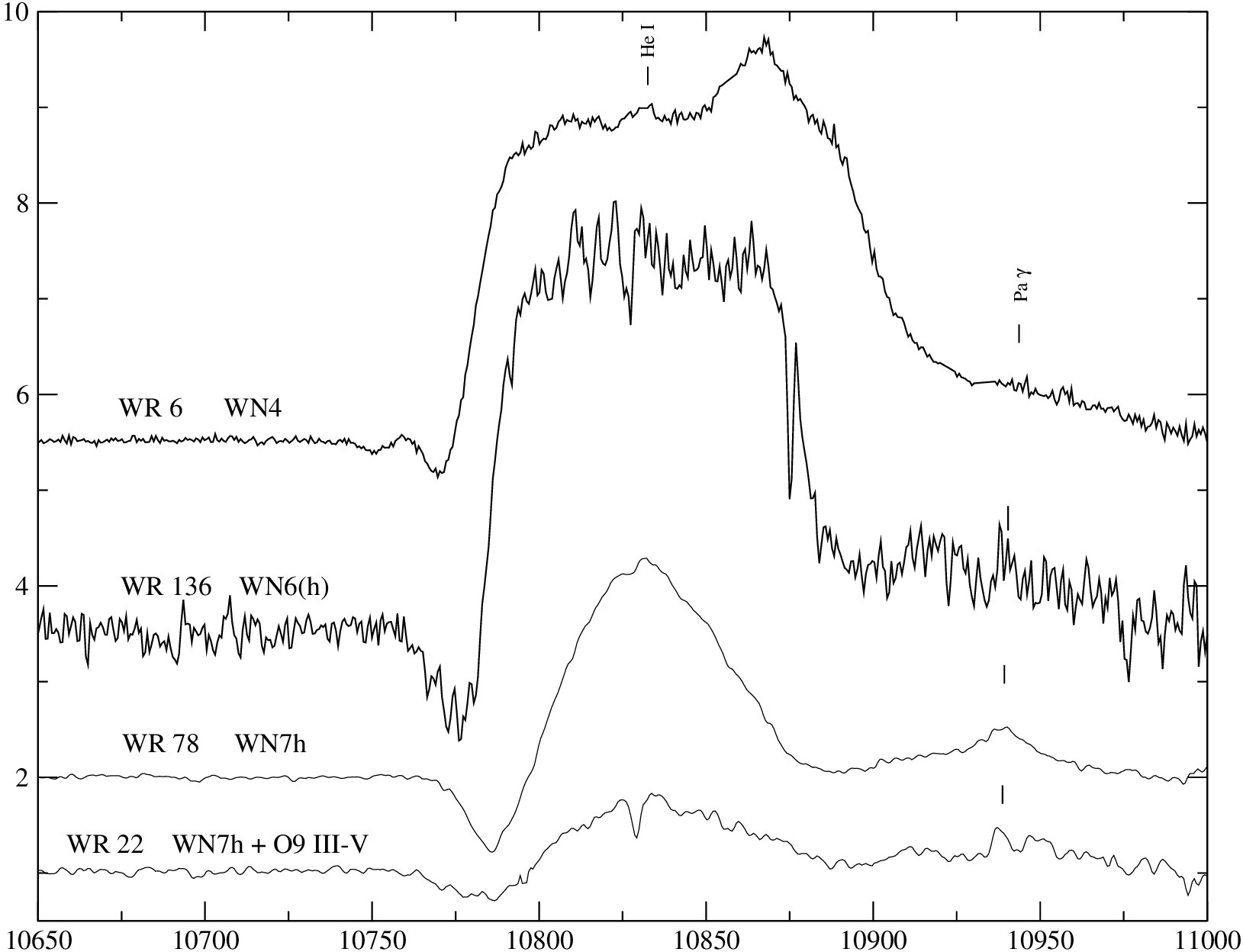}}
\caption{Spectra of the Wolf-Rayet stars from the WN subtype. From top to bottom,
WR~6, WR~136, WR~78, and WR~22. \label{wn}}
\end{figure*}

\begin{figure*}
\centering
\scalebox{0.8}{\includegraphics[width=17cm]{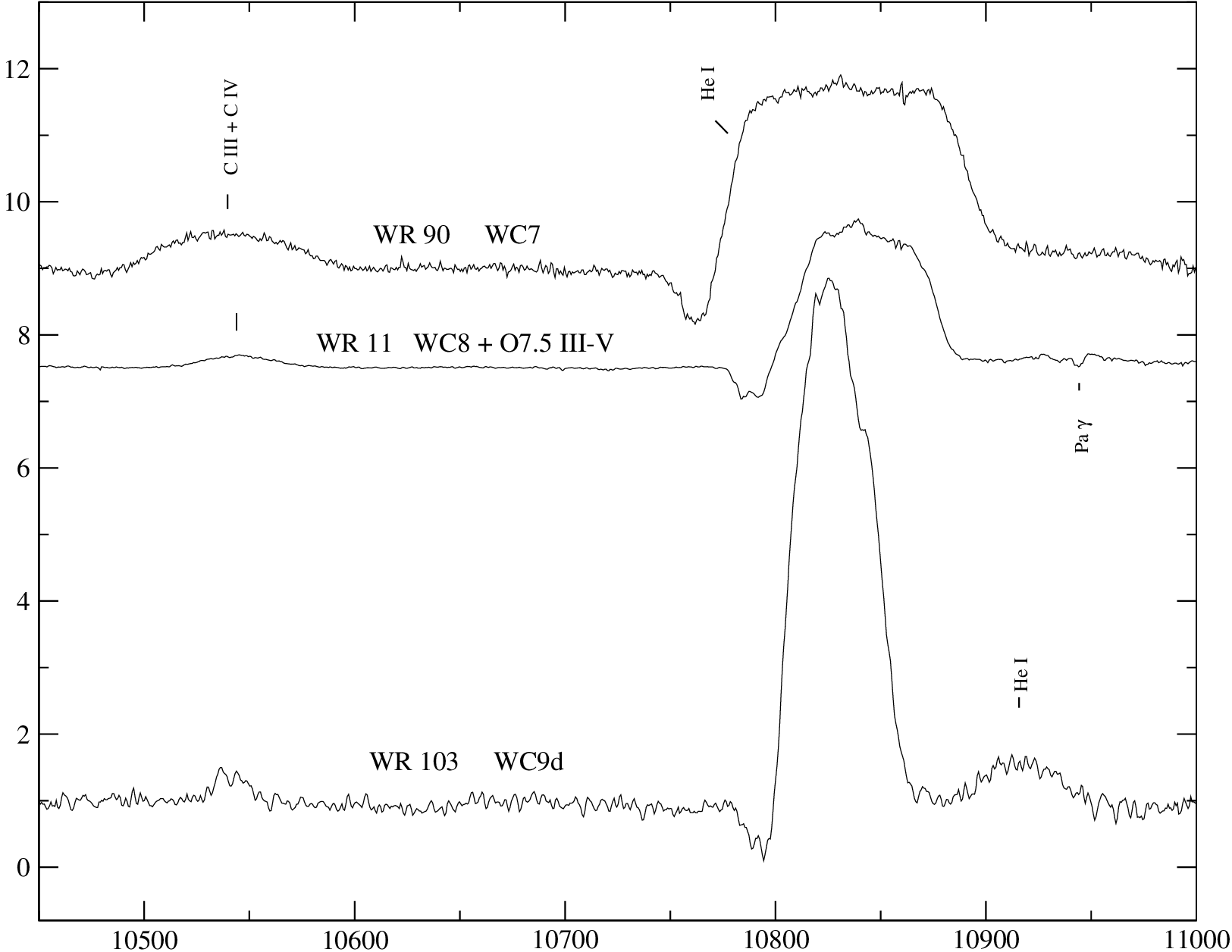}}
\caption{Same as Fig.~\ref{wn}, but for the Wolf-Rayets of the WC subtype. From top to bottom,
WR~90, WR~11 and WR~103. \label{wc}}
\end{figure*}

\begin{figure*}
\centering
\scalebox{0.8}{\includegraphics[width=17cm]{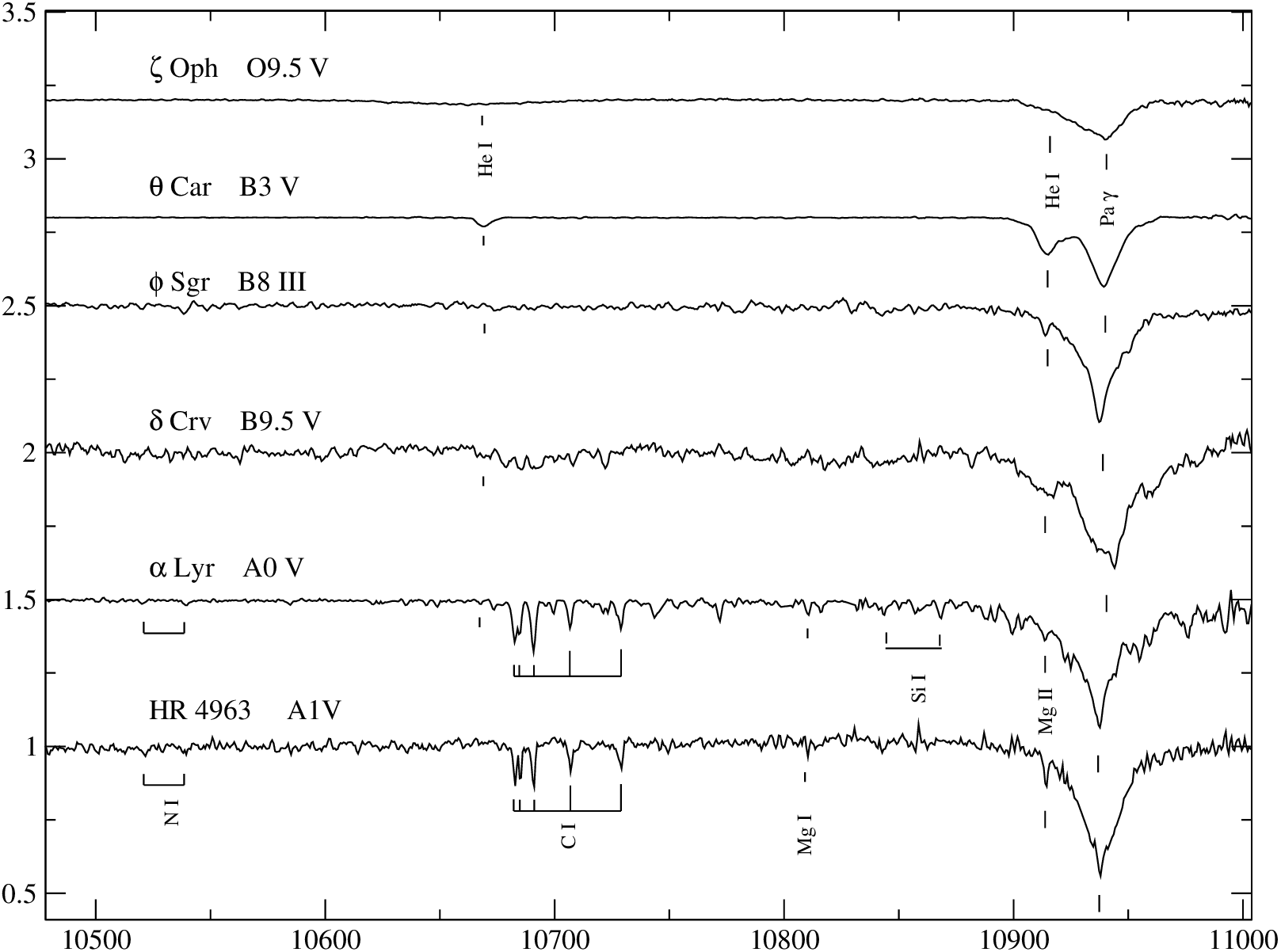}}
\caption{Spectra of non emission-line early-type stars, showing the typical photospheric lines present
in the temperature range of our sample. \label{early}}
\end{figure*}

\begin{figure*}
\centering
\scalebox{0.8}{\includegraphics[width=17cm]{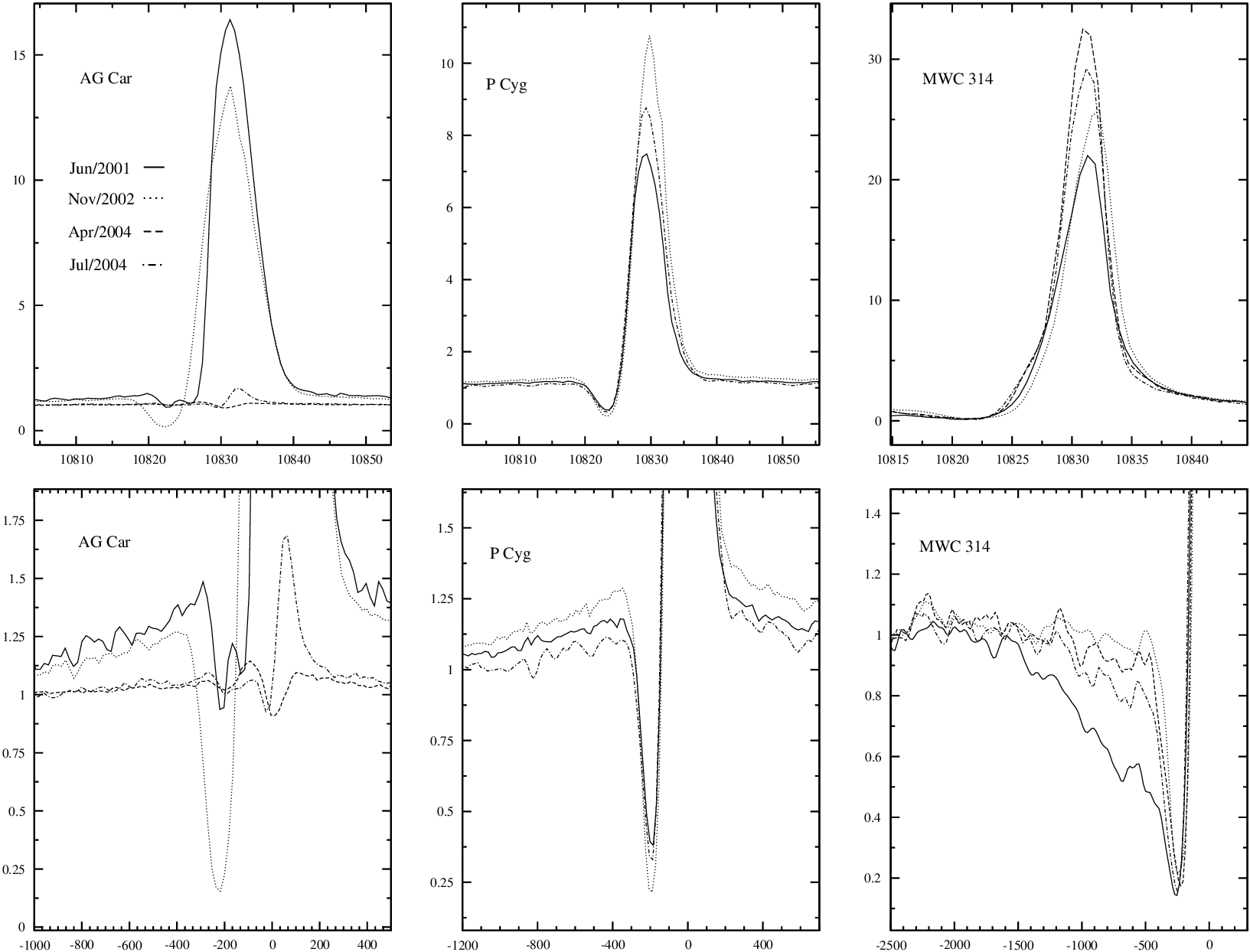}}
\caption{Spectroscopic variability of LBVs and related objects, from 2001 to 2004 (see
legend). From left to right, AG~Car, P~Cyg and MWC~314. For each star, the upper
panel presents a general view around \ion{He}{i} 10830 {\AA}, while the lower panel shows a zoom
around the P-Cyg absorption.  \label{var1}}
\end{figure*}

\begin{figure*}
\centering
\scalebox{0.8}{\includegraphics[width=17cm]{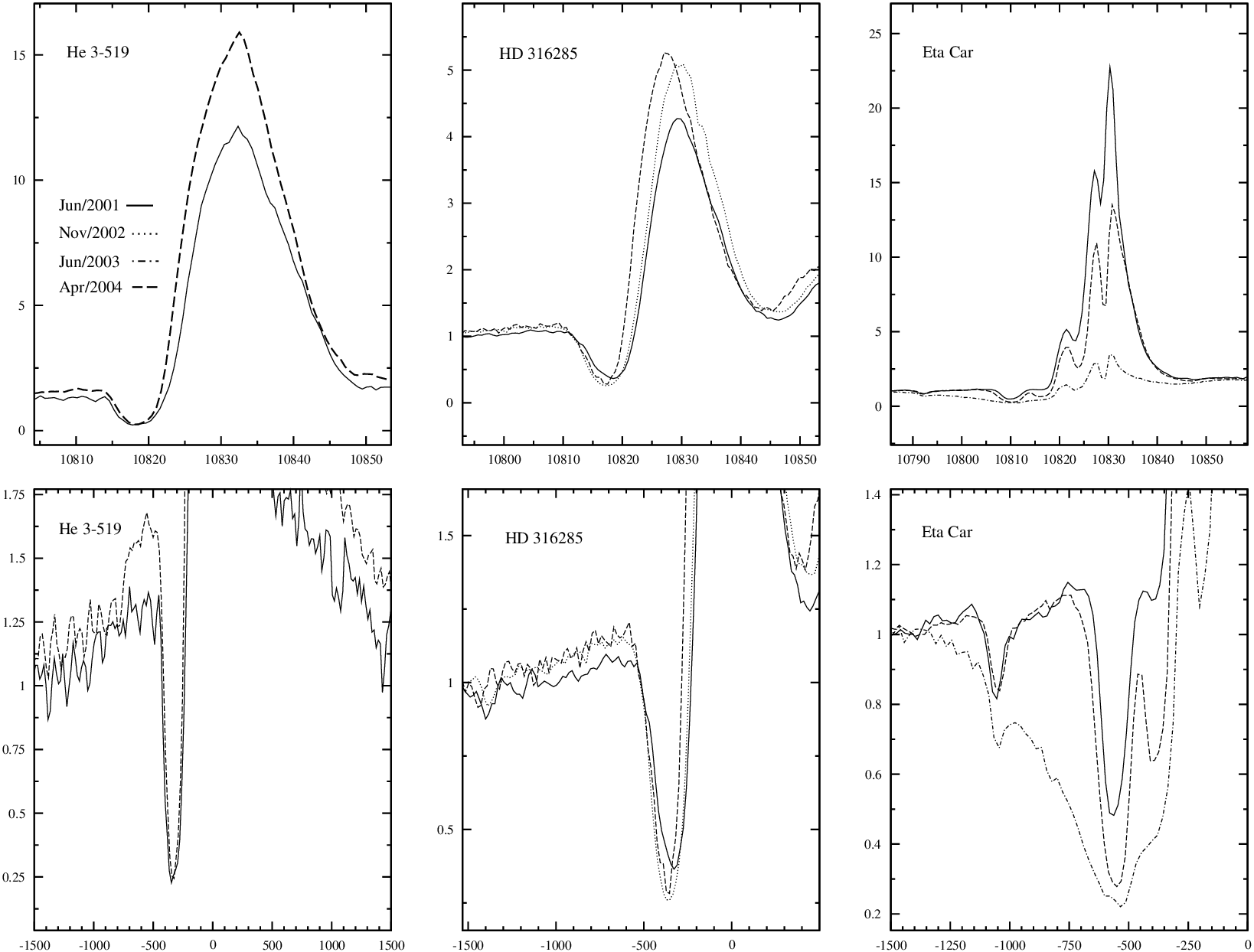}}
\caption{Same as Fig.~\ref{var1}, but for He~3-519, HD~316285 and $\eta$~Car.  \label{var2}}
\end{figure*}

\begin{figure*}
\centering
\scalebox{0.8}{\includegraphics[width=17cm]{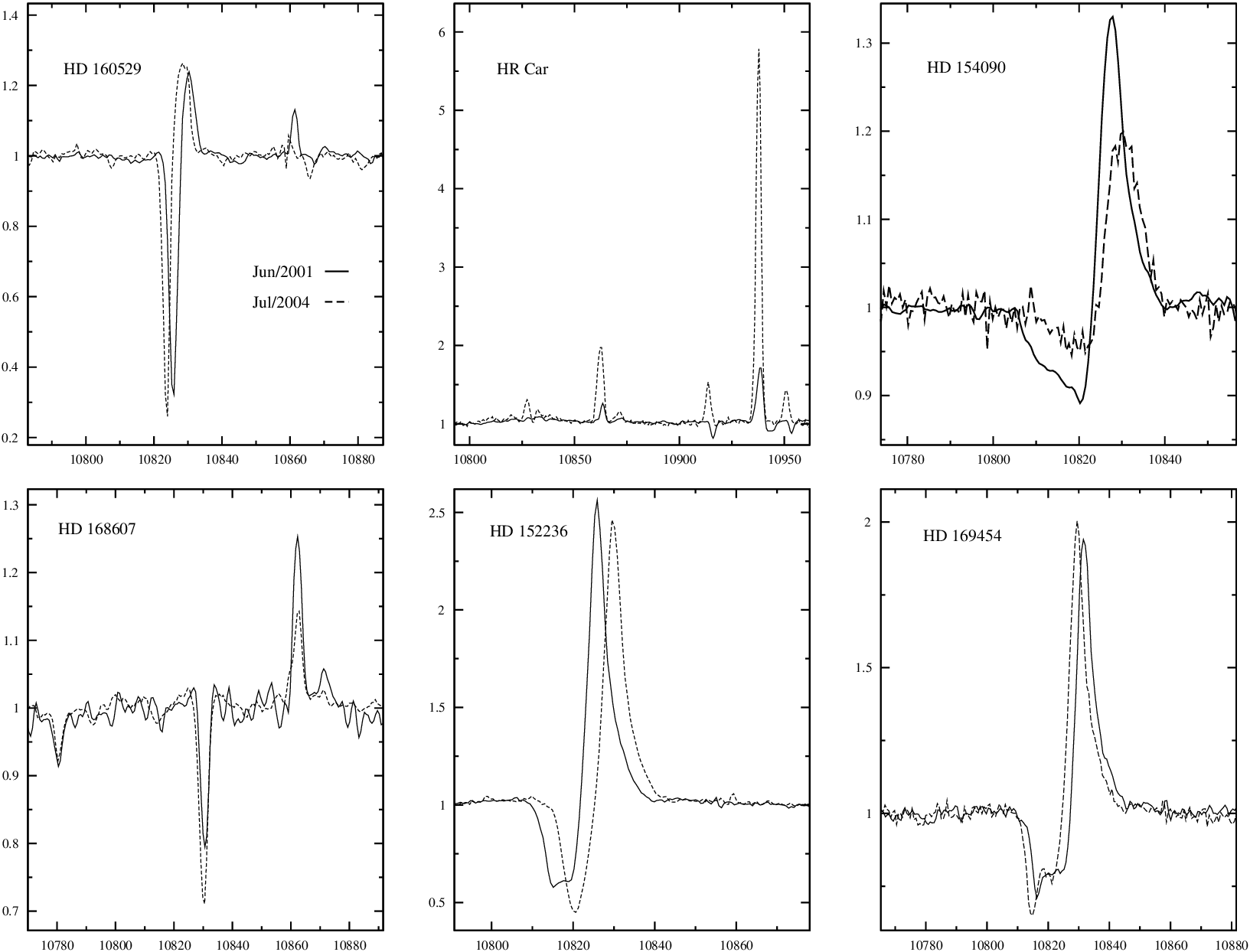}}
\caption{Same as Fig.~\ref{var1}, but for HD~160529, HR~Car, HD~154090, HD~168607, HD~152236
and HD~169454. Note the significant changes in radial velocity of HD~160529, HD~152236 and
HD~169454. \label{var3}}
\end{figure*}

However, the presence of unidentified absorption features is striking in most of the LBVs
at 10780 {\AA} and 10792 {\AA}, some of which do not have any photospheric absorption lines in
their spectra (e.g. $\eta$~Car). Figures \ref{10780a} and \ref{10780b} display a zoom around
those absorption lines. Their similar strength in stars spanning a range of spectral types,
together with their constance in variable stars such as LBVs, make it likely that the 
unidentified absorption features arise in the interstellar or circumstellar medium of those
stars. The comparison between the EW of the absorption line found at 10780 {\AA} and the color
excess E(B-V) is displayed in Figure \ref{10780_ebv}. A correlation is clearly seen
between these quantities,
\begin{equation}
\label{10780_ebv_eq}
\mathrm{E(B-V)\simeq0.16 (0.14) + 6.19 (0.53) \times EW_{10780} .}
\end{equation}

Such a tight correlation can be used to constrain circumstellar properties of LBVs. Additional
observations  with higher resolution and S/N are underway, and will be discussed in a
forthcoming paper.

Compared to the feature at 10780{\AA}, the absorption at 10792 {\AA} is weaker in
all objects except $\eta$~Car. Hence, it is difficult to perform a similar
calibration as the one made for the 10780 {\AA} feature. An observing campaign is underway
by our group to address this point. Nevertheless, this line is much stronger
in Eta Car than in other stars with similar interstellar reddening. It
was previously interpreted as a high-velocity (-1040~km\,s$^{-1}$)
component of the stellar \ion{He}{i} 10830 {\AA} \citep{damineli93,smith03}. However, this line is present
in several other LBVs which do not present any high-velocity component, and it is constant
in $\eta$~Car along its spectroscopic event. Therefore, we suggest that it is 
formed in the circumstellar environment of $\eta$~Car.
Indeed, if true, a precise calibration between EW(10792 {\AA}) and
E(B-V) may indicate the amount of circumstellar reddening present in
$\eta$~Car, which is still speculative. It would also provide constraints on
the luminosity and nature of $\eta$~Car and other objects with unknown amounts of circumstellar reddening.

\begin{acknowledgements}

We are grateful to the referee Dr. J. Puls for construtive suggestions and detailed comments. 
J. H. Groh and A. Damineli thank Brazilian
Agencies FAPESP (grants 02/11446-5 and 2005/51742-0) and CNPq (grant 200984/2004-7) for financial
support. J. H. Groh acknowledges suppport from CNPq through its undergraduate research program
(PIBIC). This work made extensive use of the SIMBAD
database, which is operated by CDS. We thank Dr.~Peter van Hoof for making available his atomic line
database.

\end{acknowledgements}

\bibliography{refs_hei10830}

\clearpage

\end{document}